\documentclass[a4paper,onecolumn,11pt,accepted=2021-12-01]{quantumarticle}
\pdfoutput=1
\usepackage[utf8]{inputenc}
\usepackage[english]{babel}
\usepackage[T1]{fontenc}
\usepackage{amsmath,amsfonts,amssymb,amsthm}
\usepackage{hyperref}

\usepackage{tikz}
\usepackage{lipsum}

\usepackage{epsfig}
\usepackage{graphicx}% Include figure files
\usepackage{dcolumn}% Align table columns on decimal point
\usepackage{bm}% bold math\input{../../../..//}
\usepackage{physics}
\usepackage{comment}
\usepackage{epstopdf}
\usepackage{dsfont}
\begin{document}

\title{Security of quantum key distribution with intensity correlations}
\author{Víctor Zapatero}
\affiliation{Escuela de Ingeniería de Telecomunicación, Department of Signal Theory and Communications, ­University of Vigo, Vigo E-36310, Spain}
\email{vzapatero@com.uvigo.es}
\author{Álvaro Navarrete}
\affiliation{Escuela de Ingeniería de Telecomunicación, Department of Signal Theory and Communications, ­University of Vigo, Vigo E-36310, Spain}
\email{anavarrete@com.uvigo.es}  
\author{Kiyoshi Tamaki}
\affiliation{Faculty of Engineering, University of Toyama, Gofuku 3190, Toyama 930-8555, Japan}
\author{Marcos Curty}
\affiliation{Escuela de Ingeniería de Telecomunicación, Department of Signal Theory and Communications, ­University of Vigo, Vigo E-36310, Spain}

\begin{abstract}
  The decoy-state method in quantum key distribution (QKD) is a popular technique to approximately achieve the performance of ideal single-photon sources by means of simpler and practical laser sources. In high-speed decoy-state QKD systems, however, intensity correlations between succeeding pulses leak information about the users’ intensity settings, thus invalidating a key assumption of this approach. Here, we solve this pressing problem by developing a general technique to incorporate arbitrary intensity correlations to the security analysis of decoy-state QKD. This technique only requires to experimentally quantify two main parameters: the correlation range and the maximum relative deviation between the selected and the actually emitted intensities. As a side contribution, we provide a non-standard derivation of the asymptotic secret key rate formula from the non-asymptotic one, in so revealing a necessary condition for the significance of the former.
\end{abstract}

\maketitle

\section{Introduction}\label{Introduction}
Quantum key distribution~\cite{Scarani,Curty,Feihu} (QKD) is a technique that enables secure and remote delivery of cryptographic keys based on the laws of quantum mechanics. The interest of QKD is that, when combined with the one-time-pad encryption scheme~\cite{OTP}, it allows for information-theoretically secure communication, unconcerned about the capabilities of future adversaries and the progress of classical or even quantum computers. For this reason, since its conception in 1984~\cite{BB84}, QKD has experienced a tremendous development both in theory and in practice, in so becoming a commercial technology that represents the most mature application of quantum information science. Nevertheless, various challenges must still be addressed in order to achieve the widespread adoption of QKD.

In real-life implementations, the information carrier of QKD is the quantum of light or photon, and due to the low transmissivity of single photons in typical optical channels ---which, for instance, in the case of optical fibers decreases exponentially with the fiber length~\cite{Pan1,Boaron,Pan2,Pan3}--- one major challenge consists of achieving high secret key generation rates at long distances. For this purpose, one natural approach is to increase the repetition rate of the laser source in the transmitter station. However, for repetition rates of the order of GHz, it has been shown that intensity correlations between succeeding pulses appear~\cite{Tomita,Geneve}, potentially opening a security loophole.

To be precise, in the absence of ideal single-photon sources, most QKD protocols use simpler laser sources that operate emitting phase-randomised weak coherent pulses (PRWCPs). This is so because PRWCPs allow the QKD users to implement the so-called decoy-state method~\cite{decoy2,decoy,decoy3,Lim}, a technique to tightly lower-bound the extractable secret key length of a QKD session. Importantly, standard decoy-state analyses rely on a fundamental assumption: for any given signal, its detection probability (or  so-called ``yield'') conditioned on the emission of a certain number of photons does not depend on the intensity of the pulse, \textit{i.e.,} on its mean photon number. Nevertheless, this assumption fails in the presence of a side channel leaking information about the intensity setting~\cite{Tamaki}, and for GHz (or higher frequency) repetition rates one such side channel is represented by intensity correlations. Intuitively, the eavesdropper (Eve) could exploit the correlations to gain information about previous intensity settings, which would allow her to make the $n$-photon yields dependent on them.

This being the case, and aiming to develop ultrafast decoy-state-based QKD systems, the question arises of how to account for arbitrary intensity correlations in the security analysis. In this regard, the existing security proofs are notably restricted. For instance, preliminary results presented in~\cite{Nagamatsu,Mizutani} deal with setting-choice-independent correlations, which neglect the possibility of information leakage and hence do not cover the major threat. On the other hand, the authors of~\cite{Tomita} go beyond setting-choice-independent correlations by providing a post-processing hardware countermeasure whose application is limited to particular instances of nearest neighbors intensity correlations. Similarly, it is worth mentioning the progress reported in~\cite{Roberts} to develop an intensity modulator (IM) that mitigates the effect of intensity correlations. Lastly, the recent work in~\cite{Pereira} presents a general technique to accomodate various other device imperfections, but it does not incorporate the use of the decoy-state method.

In this work, we provide the missing security analysis for decoy-state QKD with arbitrary intensity correlations. Despite asymptotic, our approach is experimental-friendly in the following sense. In the first place, it only requires to upper bound two parameters presumably easy to quantify in an experiment (see for instance~\cite{Geneve}): the correlation range and the maximum relative deviation between the selected intensity settings and the actually emitted intensities (which do not match in general due to the correlations). In the second place, it allows for improved secret key rates in case a specific correlations model is known to describe the IM.

As a side contribution, we use elementary results in statistical convergence to rigorously justify the asymptotic secret key rate formula from the non-asymptotic one. Crucially, the justification relies on a very natural necessary condition on the observables.
%I say natural because it is indeed something we expect: the variances must converge to 0...such that, intuitively, the variables are "asymptotically deterministic" and thus what applies to the expectations asymptotically applies to the observables too. We simply formalize this property with propositions 1 and 2.
Whenever not guaranteed by some special symmetry of the protocol (such as the delayed setting choice that enables a counterfactual scenario in the absence of information leakage), this condition must be taken as an assumption, in which case it restricts the capabilities of the adversary. In particular, the formula tolerates a restricted type of coherent attacks that we characterize in detail.
%the counterfactual setting is enabled by the fact that Eve does not see intensities but just numbers of photons. This is the "decoy-state idea" or the "delayed setting choice symmetry". Anyway, either a delayed setting choice-type symmetry applies to assure the condition, or the condition must be imposed, in which case it tolerates a restricted type of coherent attacks.

The structure of the paper goes as follows. There are six Results subsections. In the first two, Sec.~\ref{assumptions} and Sec.~\ref{main_result}, we present the general physical assumptions we impose on the intensity correlations and provide a method to quantify their effect in the parameter estimation procedure. This procedure is explained in full detail in Sec.~\ref{decoy-state} and Sec.~\ref{LP}. In Sec.~\ref{asymptotic_rate} we establish the asymptotic key rate formula and discuss the necessary condition on which it relies. The last Results subsection, Sec.~\ref{simulations}, is devoted to evaluate the rate-distance performance of our method for different values of the two parameters that characterize the correlations. In the Discussion section, Sec.~\ref{Discussion}, we summarize the contributions and limitations of our work, commenting on possible future directions. Lastly, Sec.~\ref{Cauchy}, Sec.~\ref{finite} and Sec.~\ref{propositions} are the Methods subsections, which include all the necessary technical derivations that support our results. The appendices attached at the end include a description of the channel model we use for the simulations, together with some complementary results.

%On the contrary, by giving the \texttt{accepted=YYYY-MM-DD} option, with \texttt{YYYY-MM-DD} the acceptance date, the note ``Accepted in Quantum YYYY-MM-DD, click title to verify'' can be added to the bottom of each page to clearly mark works that have been accepted in Quantum.
\section{Results}\label{Results}
For illustration purposes, we consider a standard polarization encoding decoy-state BB84 protocol~\cite{Lim}, although our results can be readily extended to any QKD protocol that relies on decoy-states. In each round $k$, with $k=1,\ldots,N$, the sender (Alice) selects a basis $x_{k}\in{}M=\{\mathrm{X,Z}\}$ with probability $q_{x_{k}}$, a uniform raw key bit $r_{k}\in\mathbb{Z}_{2}=\{0,1\}$, and an intensity setting $a_{k}\in{}A=\{\mu,\nu,\omega\}$ with probability $p_{a_{k}}$ and $\mu>\nu>\omega\geq{}0$. Note that the values of the probabilities $q_{x_{k}}$ and $p_{a_{k}}$ respectively depend on the basis and intensity settings only, but not on the particular round $k$. Then, she encodes the BB84 state defined by $x_{k}$ and $r_{k}$ in a PRWCP with intensity setting $a_{k}$, and sends it to the receiver (Bob) through the quantum channel. Importantly, as explained below, the actual mean photon number of the pulse might not match the setting $a_{k}$ due to the intensity correlations. Furthermore, regarding the transmitter, we assume perfect phase randomization, no state preparation flaws and no side-channels for simplicity. Also, possible intensity correlations in the qubit encoding (which may arise, \textit{e.g.}, when using time-bin encoding) are neglected in this work.

Similarly, Bob selects a basis $y_{k}\in{M}$ with probability $q_{y_{k}}$ (whose value, again, does not depend on the round $k$) and performs a positive operator-valued measure (POVM) on the incident pulse, given by $\{\hat{M}_{B_{k}}^{y_{k},s_{k}}\}_{s_{k}\in\{0,1,f\}}$. Here, $B_{k}$ denotes Bob's $k$-th incoming pulse, $s_{k}$ stands for Bob's classical outcome and $f$ stands for ``no click''. As usual, the basis-independent detection efficiency condition is assumed, such that $\hat{M}_{B_{k}}^{\mathrm{Z},f}=\hat{M}_{B_{k}}^{\mathrm{X},f}$. Thus, we shall simply denote these two operators by $\hat{M}_{B_{k}}^{f}$. Note that this assumption could be removed by the use of measurement-device-independent (MDI) QKD~\cite{MDI1} or twin-field QKD~\cite{TF}.
\subsection{Characterizing the intensity correlations}\label{assumptions}
Let us denote the record of intensity settings up to round $k$ by $\vec{a}_{k}=a_{1},a_{2},\ldots{},a_{k}$ ---where $a_{j}\in{}A$ for all $j$--- and let $\alpha_{k}$ denote the actual intensity delivered in round $k$. In full generality, $\alpha_{k}$ is a continuous random variable whose probability density function is fixed by the record of settings $\vec{a}_{k}$. This function, which we denote as $g_{\vec{a}_{k}}(\alpha_{k})$, is referred to as the correlations model. Below, we list three assumptions about the intensity correlations on which our analysis relies.\\

\textbf{Assumption 1.} As supported by GHz-clock QKD experiments~\cite{Tomita,Geneve}, we shall consider that the correlations do not compromise the poissonian character of the photon number statistics of the source conditioned on the value of the actual intensity, $\alpha_{k}$. That is to say, for any given round $k$, and for all $n_{k}\in\mathbb{N}$,
\begin{equation}\label{statistics_conditional}
	p(n_{k}|\alpha_{k})=\frac{e^{-\alpha_{k}}\alpha_{k}^{n_{k}}}{n_{k}!}.
\end{equation}
\\

\textbf{Assumption 2.} For all possible records $\vec{a}_{k}$, we shall assume that
\begin{equation}\label{fluctuation}
	\abs{1-\frac{\alpha_{k}}{a_{k}}}\leq{}\delta_{\rm max}.
\end{equation}
%CRUCIAL: this assumption is physical and testable
Namely, for every round, we impose that $g_{\vec{a}_{k}}(\alpha_{k})$ is only nonzero for $\alpha_{k}\in[a_{k}^{-},a_{k}^{+}]$, where $a_{k}^{\pm}=a_{k}(1\pm{\delta_{\rm max}})$. Thus, $\delta_{\rm max}$ defines the maximum relative deviation between $a_{k}$ and $\alpha_{k}$. Note that we are assuming here that the value of $\delta_{\rm max}$ does not depend on $a_{k}$ for simplicity, but such dependence could be easily incorporated in the analysis to obtain slightly tighter results. Also, we remark that a bound of the type of Eq.~(\ref{fluctuation}) has been quantified in a recent experiment reported in~\cite{Geneve}.

From Eq.~(\ref{statistics_conditional}) and Eq.~(\ref{fluctuation}), it follows that the photon number statistics of round $k$ read
\begin{equation}\label{statistics} p_{n_{k}}\lvert_{\vec{a}_{k}}=\int_{a_{k}^{-}}^{a_{k}^{+}}g_{\vec{a}_{k}}(\alpha_{k})\frac{e^{-\alpha_{k}}\alpha_{k}^{n_{k}}}{n_{k}!}\,d\alpha_{k},
\end{equation}
for all $n_{k}\in\mathbb{N}$.\\

\textbf{Assumption 3.} We assume that the intensity correlations have a finite range, say $\xi$, meaning that $g_{\vec{a}_{k}}(\alpha_{k})$ is independent of those previous settings $a_{j}$ with $k-j>\xi$.\\

Beyond the assumptions presented here, we shall consider that $g_{\vec{a}_{k}}(\alpha_{k})$ is unknown, such that our results are model-independent.
\subsection{Quantifying the effect of the intensity correlations}\label{main_result}

Here, we rely on the three assumptions introduced in Sec.~\ref{assumptions} to account for the effect of intensity correlations in the decoy-state analysis. A key idea ---originally presented in~\cite{Tamaki} to deal with Trojan horse attacks--- is to pose a restriction on the maximum bias that Eve can induce between the $n$-photon yields associated to different intensity settings. For this purpose, we use a fundamental result presented in~\cite{Lo-Preskill} and further developed in~\cite{Pereira}. Since this result is a direct consequence of the Cauchy–Schwarz (CS) inequality in complex Hilbert spaces, we shall refer to it as the CS constraint. The reader is referred to the Methods Sec.~\ref{Cauchy} for a definition of this result, and below we present the relevant restrictions we derive with it.

Precisely, for any given round $k$, photon number $n\in\mathbb{N}$, intensity setting $c\in{}A$ and bit value $r\in\{0,1\}$, we define the yield $Y_{n,c}^{(k)}=p^{(k)}(\mathrm{click}|n,c,\mathrm{Z,Z})$ and the error probability $H_{n,c,r}^{(k)}=p^{(k)}(\mathrm{err}|n,c,\mathrm{X,X},r)$, where the right-hand sides are shorthand for $p(s_{k}\neq{}f|n_{k}=n,a_{k}=c,x_{k}=\mathrm{Z},y_{k}=\mathrm{Z})$ and $p(s_{k}\neq{}f,s_{k}\neq{}r_{k}|n_{k}=n,a_{k}=c,x_{k}=\mathrm{X},y_{k}=\mathrm{X},r_{k}=r)$, respectively. Then, a major result of this work is to show that, for any two distinct intensity settings $a$ and $b$, their yields and error probabilities satisfy
\begin{eqnarray}\label{deviation}
	&&G_{-}\left(Y_{n,a}^{(k)},\tau_{ab,n}^{\xi}\right)\leq{}Y_{n,b}^{(k)}\leq{}G_{+}\left(Y_{n,a}^{(k)},\tau_{ab,n}^{\xi}\right),\nonumber \\
	&&G_{-}\left(H_{n,a,r}^{(k)},\tau_{ab,n}^{\xi}\right)\leq{}H_{n,b,r}^{(k)}\leq{}G_{+}\left(H_{n,a,r}^{(k)},\tau_{ab,n}^{\xi}\right),
\end{eqnarray}
for all $k$, $n$ and $r$, where
\begin{equation}\label{Gs}
	G_{-}(y,z) = \begin{cases}
		g_{-}(y,z) & \mathrm{if}\hspace{.3cm}y>1-z \\
		0 & \mathrm{otherwise}  
	\end{cases}
	\hspace{.5cm}\mathrm{and}\hspace{.5cm}
	G_{+}(y,z) = \begin{cases}
		g_{+}(y,z) & \mathrm{if}\hspace{.3cm}y<z \\
		1 & \mathrm{otherwise}  
	\end{cases}
\end{equation}
with $g_{\pm}(y,z)=y+(1-z)(1-2y)\pm{}2\sqrt{z(1-z)y(1-y)}$, $\xi$ stands for the correlation range and
\begin{equation}\label{tau}
	\tau_{ab,n}^{\xi}= \begin{cases}
		e^{a^{-}+b^{-}-(a^{+}+b^{+})}\left[1-\displaystyle\sum_{c \in A}p_{c}\left(e^{-c^{-}}-e^{-c^{+}}\right)\right]^{2\xi} & \mathrm{if}\hspace{.3cm}n=0,\\
		e^{a^{+}+b^{+}-(a^{-}+b^{-})}\left(\frac{a^{-}b^{-}}{a^{+}b^{+}}\right)^{n}\left[1-\displaystyle\sum_{c \in A}p_{c}\left(e^{-c^{-}}-e^{-c^{+}}\right)\right]^{2\xi} & \mathrm{if}\hspace{.3cm}n\geq{}1. 
	\end{cases}
\end{equation}
The full derivation of this result is given in the Methods Sec.~\ref{Cauchy}.

Notably, Eq.~(\ref{deviation}) must be combined with a decoy-state method in order to estimate the numbers of counts and errors triggered by single photon emissions, which determine the secret key rate. For this purpose, in Sec.~\ref{decoy-state} we provide a decoy-state analysis that relies on assumptions 1 and 2
%the finite range hypothesis is not used in the decoy-state analysis!
to deal with intensity correlations, and in Sec.~\ref{LP} we present the resulting linear programs that fulfill the parameter estimation. In this regard, since the constraints of Eq.~(\ref{deviation}) are non-linear, first-order approximations with respect to some reference parameters are derived from them, which we refer to as the linearized CS constraints. Importantly, replacing the original functions by their linear approximations leads to looser but valid constraints too, thanks to the convexity of these functions.
%Indeed, from the point of view of the decoy-state analysis, the only relevant feature is the fluctuation of the intensities, and not the correlations themselves.
\subsection{Decoy-state method}\label{decoy-state}
The Z basis gain with intensity setting $a$ is defined as $Z_{a,N}=\sum_{k=1}^{N}Z_{a}^{(k)}$ with $Z_{a}^{(k)}=\mathds{1}_{\{a_{k}=a,x_{k}=y_{k}=\mathrm{Z},s_{k}\neq{}f\}}$. That is to say, $Z_{a}^{(k)}=1$ if, in round $k$, both parties select the Z basis, Alice selects intensity setting $a$ and a click occurs, and zero otherwise. Thus,
\begin{equation}\label{gain}
	\bigl\langle{Z_{a}^{(k)}}\bigr\rangle=p^{(k)}(a,\mathrm{Z,Z,click})=q_{\rm Z}^{2}p_{a}\sum_{n=0}^{\infty}p^{(k)}(n,\mathrm{click}|a,\mathrm{Z,Z})=q_{\rm Z}^{2}p_{a}\sum_{n=0}^{\infty}p^{(k)}(n|a)Y_{n,a}^{(k)},
\end{equation}
where the yield $Y_{n,a}^{(k)}$ is defined in Sec.~\ref{main_result} and one can generically refer to the $\langle{Z_{a}^{(k)}}\rangle$ as the ``expected gains of round $k$''. Going back to Eq.~(\ref{gain}), note that
\begin{equation}\label{un-referenced}
p^{(k)}(n|a)=\sum_{\vec{a}_{k-1}}p_{a_{1}}\ldots{}p_{a_{k-1}}p_{n}\lvert_{a,\vec{a}_{k-1}},
\end{equation}
and in virtue of Eq.~(\ref{statistics}), the record-independent bounds
\begin{equation}\label{intervals}
	p^{(k)}(0|a)\in{}\left[e^{-a^{+}},e^{-a^{-}}\right]\hspace{.4cm}\mathrm{and}\hspace{.4cm}p^{(k)}(n|a)\in{}\left[\frac{e^{-a^{-}}a^{-\hspace{.06cm}n}}{n!},\frac{e^{-a^{+}}a^{+\hspace{.06cm}n}}{n!}\right]\hspace{.1cm}(n\geq{}1)
\end{equation}
follow from the decreasing (increasing) character of $e^{-x}x^{n}$ for $n=0$ ($n\geq{}1$) in the interval $x\in(0,1)$. Explicitly using these intervals in Eq.~(\ref{gain}), one obtains
\begin{equation}\label{yields}
	\frac{\bigl\langle{Z_{a}^{(k)}}\bigr\rangle}{q_{\rm Z}^{2}p_{a}}\geq{}e^{-a^{+}}Y_{0,a}^{(k)}+\sum_{n=1}^{\infty}\frac{e^{-a^{-}}a^{-\hspace{.06cm}n}}{n!}Y_{n,a}^{(k)}\hspace{.4cm}\mathrm{and}\hspace{.4cm}\frac{\bigl\langle{Z_{a}^{(k)}}\bigr\rangle}{q_{\rm Z}^{2}p_{a}}\leq{}e^{-a^{-}}Y_{0,a}^{(k)}+\sum_{n=1}^{\infty}\frac{e^{-a^{+}}a^{+\hspace{.06cm}n}}{n!}Y_{n,a}^{(k)}
\end{equation}
for all $a\in{}A$ and $k=1,\ldots,N$. Further selecting a threshold photon number for the numerics, $n_{\rm cut}$, from Eq.~(\ref{yields}) we have
\begin{eqnarray}\label{yields_2}
	&&\frac{\bigl\langle{Z_{a}^{(k)}}\bigr\rangle}{q_{\rm Z}^{2}p_{a}}\geq{}e^{-a^{+}}Y_{0,a}^{(k)}+\sum_{n=1}^{n_{\rm cut}}\frac{e^{-a^{-}}a^{-\hspace{.06cm}n}}{n!}Y_{n,a}^{(k)}\hspace{.4cm}\mathrm{and}\nonumber \\
	&&\frac{\bigl\langle{Z_{a}^{(k)}}\bigr\rangle}{q_{\rm Z}^{2}p_{a}}\leq{}1-e^{-a^{+}}+e^{-a^{-}}Y_{0,a}^{(k)}-\sum_{n=1}^{n_{\rm cut}}\frac{e^{-a^{+}}a^{+\hspace{.06cm}n}}{n!}\left(1-Y_{n,a}^{(k)}\right)
\end{eqnarray}
for all $a\in{}A$ and $k=1,\ldots,N$, where in the second inequality we have used the fact that $\sum_{n=n_{\rm cut}+1}^{\infty}Y_{n,a}^{(k)}e^{-a^{+}}a^{+\hspace{.06cm}n}/n!\leq{}1-\sum_{n=0}^{n_{\rm cut}}e^{-a^{+}}a^{+\hspace{.06cm}n}/n!$. Importantly, replacing Z by X everywhere, one obtains the corresponding analysis for the X basis gains and yields of round $k$.\\

On the other hand, similar constraints can be imposed on the error counts. To be precise, the number of X basis error counts with setting $a$ is defined as $E_{a}=\sum_{k=1}^{N}E_{a}^{(k)}$ with $E_{a}^{(k)}=X_{a}^{(k)}\mathds{1}_{\{r_{k}\neq{}s_{k}\}}$, such that
\begin{equation}
	\bigl\langle{E_{a}^{(k)}}\bigr\rangle=p^{(k)}(a,\mathrm{X,X,err})=q_{\rm X}^{2}p_{a}\sum_{n=0}^{\infty}p^{(k)}(n,\mathrm{err}|a,\mathrm{X,X})=q_{\rm X}^{2}p_{a}\sum_{n=0}^{\infty}p^{(k)}(n|a)H_{n,a}^{(k)},
\end{equation}
where we defined $H_{n,a}^{(k)}=p^{(k)}(\mathrm{err}|n,a,\mathrm{X,X})=(H_{n,a,0}^{(k)}+H_{n,a,1}^{(k)})/2$. Now, making use of Eq.~(\ref{intervals}) as before and selecting a threshold photon number $n_{\rm cut}$, it follows that
\begin{eqnarray}\label{errors}
	&&\frac{\bigl\langle{E_{a}^{(k)}}\bigr\rangle}{q_{\rm X}^{2}p_{a}}\geq{}e^{-a^{+}}H_{0,a}^{(k)}+\sum_{n=1}^{n_{\rm cut}}\frac{e^{-a^{-}}a^{-\hspace{.06cm}n}}{n!}H_{n,a}^{(k)}\hspace{.4cm}\mathrm{and}\nonumber \\
	&&\frac{\bigl\langle{E_{a}^{(k)}}\bigr\rangle}{q_{\rm X}^{2}p_{a}}\leq{}1-e^{-a^{+}}+e^{-a^{-}}H_{0,a}^{(k)}-\sum_{n=1}^{n_{\rm cut}}\frac{e^{-a^{+}}a^{+\hspace{.06cm}n}}{n!}\left(1-H_{n,a}^{(k)}\right)
\end{eqnarray}
for all $a\in{}A$ and $k=1,\ldots,N$.

At this point, summing over $k$ and dividing by $N$ both in Eq.~(\ref{yields_2}) and Eq.~(\ref{errors}), one trivially obtains bounds for the average parameters $y_{n,a,N}=\sum_{k=1}^{N}Y_{n,a}^{(k)}/N$ and $h_{n,a,N}=\sum_{k=1}^{N}H_{n,a}^{(k)}/N$ from the round-dependent bounds. Namely, defining $\overline{Z}_{a,N}=Z_{a,N}/N$ and $\overline{E}_{a,N}=E_{a,N}/N$, the final bounds are
\begin{eqnarray}\label{constraints_decoy}
	&&\frac{\left\langle{\overline{Z}_{a,N}}\right\rangle}{q_{\rm Z}^{2}p_{a}}\geq{}e^{-a^{+}}y_{0,a}+\sum_{n=1}^{n_{\rm cut}}\frac{e^{-a^{-}}a^{-\hspace{.06cm}n}}{n!}y_{n,a,N},\nonumber \\
	&&\frac{\left\langle{\overline{Z}_{a,N}}\right\rangle}{q_{\rm Z}^{2}p_{a}}\leq  1-e^{-a^{+}}+e^{-a^{-}}y_{0,a}-\sum_{n=1}^{n_{\rm cut}}\frac{e^{-a^{+}}a^{+\hspace{.06cm}n}}{n!}\left(1-y_{n,a,N}\right)\hspace{.4cm}\mathrm{and}\nonumber \\
	&&\frac{\left\langle{\overline{E}_{a,N}}\right\rangle}{q_{\rm X}^{2}p_{a}}\geq  e^{-a^{+}}h_{0,a}+\sum_{n=1}^{n_{\rm cut}}\frac{e^{-a^{-}}a^{-\hspace{.06cm}n}}{n!}h_{n,a,N},\nonumber \\
	&&\frac{\left\langle{\overline{E}_{a,N}}\right\rangle}{q_{\rm X}^{2}p_{a}}\leq  1-e^{-a^{+}}+e^{-a^{-}}h_{0,a}-\sum_{n=1}^{n_{\rm cut}}\frac{e^{-a^{+}}a^{+\hspace{.06cm}n}}{n!}\left(1-h_{n,a,N}\right),
\end{eqnarray}
for a common threshold photon number $n_{\rm cut}$ and for all $a\in{}A$.
\subsection{Linear programs for parameter estimation}\label{LP}
Even though Eq.~(\ref{deviation}) provides the relevant restrictions on the maximum bias that Eve can induce between different yields/error probabilities, it consists of a set of non-linear constraints unsuitable for parameter estimation via linear programming. As mentioned in Sec.~\ref{main_result} though, in virtue of the convexity/concavity of the functions that define the constraints, their first-order expansions around any given reference yield/error provide valid linear bounds as well. For instance, if we focus on the yields, we have that $G_{-}(Y_{n,a}^{(k)},\tau_{ab,n}^{\xi})\geq{}G_{-}(\tilde{Y}_{n,a}^{(k)},\tau_{ab,n}^{\xi})+{G'}_{-}(\tilde{Y}_{n,a}^{(k)},\tau_{ab,n})(Y_{n,a}^{(k)}-\tilde{Y}_{n,a}^{(k)})$ and $G_{+}(Y_{n,a}^{(k)},\tau_{ab,n}^{\xi})\leq{}G_{+}(\tilde{Y}_{n,a}^{(k)},\tau_{ab,n}^{\xi})+{G'}_{+}(\tilde{Y}_{n,a}^{(k)},\tau_{ab,n}^{\xi})(Y_{n,a}^{(k)}-\tilde{Y}_{n,a}^{(k)})$ for all $Y_{n,a}^{(k)}\in(0,1)$, irrespectively of which reference yields $\tilde{Y}_{n,a}^{(k)}\in(0,1)$ we select for the linear expansion. Also, note that the derivative functions ${G'}_{\pm}$ are well-defined for all $Y_{n,a}^{(k)}\in(0,1)$ because the $G_{\pm}$ are smooth piecewise functions. In particular,
\begin{equation}\label{Kato_derived}
	{G'}_{-}(y,z) = \begin{cases}
		{g'}_{-}(y,z) & \mathrm{if}\hspace{.3cm}y>1-z \\
		0 & \mathrm{otherwise}  
	\end{cases}
	\hspace{.5cm}\mathrm{and}\hspace{.5cm}
	{G'}_{+}(y,z) = \begin{cases}
		{g'}_{+}(y,z) & \mathrm{if}\hspace{.3cm}y<z \\
		0 & \mathrm{otherwise}  
	\end{cases}
\end{equation}
with ${g'}_{\pm}(y,z)=-1+2z\pm(1-2y)\sqrt{{z(1-z)}/{y(1-y)}}$. Thus, given a reference yield $\tilde{Y}_{n,a}^{(k)}\in(0,1)$, the linearized bounds are
\begin{eqnarray}\label{KATO_1}
	&&G_{-}\left(\tilde{Y}_{n,a}^{(k)},\tau_{ab,n}^{\xi}\right)+{G'}_{-}\left(\tilde{Y}_{n,a}^{(k)},\tau_{ab,n}^{\xi}\right)\left(Y_{n,a}^{(k)}-\tilde{Y}_{n,a}^{(k)}\right)\leq{}{Y}_{n,b}^{(k)}\leq{}\nonumber \\
	&&G_{+}\left(\tilde{Y}_{n,a}^{(k)},\tau_{ab,n}^{\xi}\right)+{G'}_{+}\left(\tilde{Y}_{n,a}^{(k)},\tau_{ab,n}^{\xi}\right)\left(Y_{n,a}^{(k)}-\tilde{Y}_{n,a}^{(k)}\right).
\end{eqnarray}

Identically, for any given reference $n$-photon error click probabilities $\tilde{H}_{n,a,r}^{(k)}\in(0,1)$, the linearized versions of the corresponding constraints read
\begin{eqnarray}\label{KATO_2_r}
	&&G_{-}\left(\tilde{H}_{n,a,r}^{(k)},\tau_{ab,n}^{\xi}\right)+{G'}_{-}\left(\tilde{H}_{n,a,r}^{(k)},\tau_{ab,n}^{\xi}\right)\left(H_{n,a,r}^{(k)}-\tilde{H}_{n,a,r}^{(k)}\right)\leq{}{H}_{n,b,r}^{(k)}\leq{}\nonumber \\
	&&G_{+}\left(\tilde{H}_{n,a,r}^{(k)},\tau_{ab,n}^{\xi}\right)+{G'}_{+}\left(\tilde{H}_{n,a,r}^{(k)},\tau_{ab,n}^{\xi}\right)\left(H_{n,a,r}^{(k)}-\tilde{H}_{n,a,r}^{(k)}\right).
\end{eqnarray}
If, in addition, we select reference parameters independent of $r$, say $\tilde{H}_{n,a,r}^{(k)}=\tilde{H}_{n,a}^{(k)}$ for both $r=0$ and $r=1$, the linearized lower (upper) bound has the exact same slope and the exact same intercept for both $r=0$ and $r=1$. As a consequence, the relevant error probabilities entering the decoy-state analysis, $H_{n,a}^{(k)}=(H_{n,a,0}^{(k)}+H_{n,a,1}^{(k)})/2$ and $H_{n,b}^{(k)}=(H_{n,b,0}^{(k)}+H_{n,b,1}^{(k)})/2$, trivially verify
\begin{eqnarray}\label{KATO_2}
	&&G_{-}\left(\tilde{H}_{n,a}^{(k)},\tau_{ab,n}^{\xi}\right)+{G'}_{-}\left(\tilde{H}_{n,a}^{(k)},\tau_{ab,n}^{\xi}\right)\left(H_{n,a}^{(k)}-\tilde{H}_{n,a}^{(k)}\right)\leq{}{H}_{n,b}^{(k)}\leq{}\nonumber \\
	&&G_{+}\left(\tilde{H}_{n,a}^{(k)},\tau_{ab,n}^{\xi}\right)+{G'}_{+}\left(\tilde{H}_{n,a}^{(k)},\tau_{ab,n}^{\xi}\right)\left(H_{n,a}^{(k)}-\tilde{H}_{n,a}^{(k)}\right),
\end{eqnarray}
as we wanted to show.

As a final comment, note that, for all practical purposes, one can restrict the reference parameters to be round-independent: $\tilde{Y}_{n,a}^{(k)}=\tilde{y}_{n,a}$ and $\tilde{H}_{n,a}^{(k)}=\tilde{h}_{n,a}$ for all $k=1,\ldots,N$. This being the case, summing over $k$ and dividing by $N$ in Eq.~(\ref{KATO_1}) and Eq.~(\ref{KATO_2}), one obtains respective inequalities for the average parameters $y_{n,b,N}=\sum_{k=1}^{N}Y_{n,b}^{(k)}/N$ and $h_{n,b,N}=\sum_{k=1}^{N}H_{n,b}^{(k)}/N$. Namely, for all $a\in{}A$, $b\in{}A$ ($b\neq{}a$) and $n\in{}\mathbb{N}$, we have $c^{-}_{ab,n}+m^{-}_{ab,n}y_{n,a,N}\leq{}{y}_{n,b,N}\leq{}c^{+}_{ab,n}+m^{+}_{ab,n}y_{n,a,N}$ and $t^{-}_{ab,n}+s^{-}_{ab,n}h_{n,a,N}\leq{}{h}_{n,b,N}\leq{}t^{+}_{ab,n}+s^{+}_{ab,n}h_{n,a,N}$, where, for conciseness, we define the intercepts and slopes
\begin{eqnarray}\label{intercepts} &&c^{\pm}_{ab,n}=G_{\pm}(\tilde{y}_{n,a},\tau_{ab,n}^{\xi})-{G'}_{\pm}(\tilde{y}_{n,a},\tau_{ab,n}^{\xi})\tilde{y}_{n,a},\hspace{.2cm}m^{\pm}_{ab,n}={G'}_{\pm}(\tilde{y}_{n,a},\tau_{ab,n}^{\xi}),\nonumber \\
	&&t^{\pm}_{ab,n}=G_{\pm}(\tilde{h}_{n,a},\tau_{ab,n}^{\xi})-{G'}_{\pm}(\tilde{h}_{n,a},\tau_{ab,n}^{\xi})\tilde{h}_{n,a}\hspace{.2cm}\mathrm{and}\hspace{.2cm}s^{\pm}_{ab,n}={G'}_{\pm}(\tilde{h}_{n,a},\tau_{ab,n}^{\xi}).
\end{eqnarray}
Of course, the tightness of these linear bounds is subject to the adequacy of the selected reference yields, and thus it relies on a characterization of the quantum channel. Note, however, that aiming to further improve the results, one could incorporate more linearized CS constraints to the problem by using various reference yields for each pair $(n,a)$, instead of just one. Also, we recall that simpler bounds not relying on any reference values can be derived by using the so-called trace distance argument~\cite{Nielsen}, and this is what we do in Appendix~\ref{TD}. Another alternative would be, of course, to solve a non-linear optimization problem given by the original CS constraints.\\

To finish with, we present the linear programs that allow to estimate the relevant single-photon parameters, putting together the decoy-state constraints introduced in Sec.~\ref{decoy-state} and the above linearized CS constraints. In the first place, we address the average number of signal-setting single-photon counts, defined as $\overline{Z}_{1,\mu,N}=\sum_{k=1}^{N}Z_{1,\mu}^{(k)}/N$ with $Z_{1,\mu}^{(k)}=Z_{\mu}^{(k)}\mathds{1}_{\{n_{k}=1\}}$. Since $\bigl\langle{Z_{1,\mu}^{(k)}}\bigr\rangle=q_{\rm Z}^{2}p_{\mu}p^{(k)}(1|\mu)Y_{1,\mu}^{(k)}\geq{}q_{\rm Z}^{2}p_{\mu}{\mu}^{-}e^{-\mu^{-}}Y_{1,\mu}^{(k)}$, averaging over $k$ it follows that
\begin{equation}
	\langle{\overline{Z}_{1,\mu,N}}\rangle\geq{}q_{\rm Z}^{2}p_{\mu}{\mu}^{-}e^{-\mu^{-}}y_{1,\mu,N},
\end{equation}
and a lower bound $y_{1,\mu,N}^{\rm L}$ on $y_{1,\mu,N}$ is reached by the following linear program:
\begin{gather}\label{lp_1}
	\min\quad y_{1,\mu,N} \\
	\begin{aligned}
		\textup{s.t.}\hspace{.3cm}&\frac{\left\langle{\overline{Z}_{a,N}}\right\rangle}{q_{\rm Z}^{2}p_{a}}\geq  e^{-a^{+}}y_{0,a}+\sum_{n=1}^{n_{\rm cut}}\frac{e^{-a^{-}}a^{-\hspace{.06cm}n}}{n!}y_{n,a,N}\hspace{.2cm}(a\in{}A),\nonumber\\
		&\frac{\left\langle{\overline{Z}_{a,N}}\right\rangle}{q_{\rm Z}^{2}p_{a}}\leq  1-e^{-a^{+}}+e^{-a^{-}}y_{0,a}-\sum_{n=1}^{n_{\rm cut}}\frac{e^{-a^{+}}a^{+\hspace{.06cm}n}}{n!}\left(1-y_{n,a,N}\right)\hspace{.2cm}(a\in{}A),\nonumber\\
		&c^{+}_{ab,n}+m^{+}_{ab,n}y_{n,a,N}\geq{}{y}_{n,b,N}\hspace{.2cm}(a\in{}A,\ b\in{}A,\ b\neq{}a,\ n=0,\ldots{},n_{\rm cut}),\nonumber\\
		&c^{-}_{ab,n}+m^{-}_{ab,n}y_{n,a,N}\leq{}{y}_{n,b,N}\hspace{.2cm}(a\in{}A,\ b\in{}A,\ b\neq{}a,\ n=0,\ldots{},n_{\rm cut}),\nonumber\\
		&0\leq{}y_{n,a,N}\leq{}1\hspace{.2cm}(a\in{}A,n=0,\ldots,n_{\rm cut}).
	\end{aligned}
\end{gather}
We recall that the $c^{\pm}_{ab,n}$ and the $m^{\pm}_{ab,n}$ are defined in Eq.~(\ref{intercepts}). Needless to say, replacing Z by X everywhere one obtains the corresponding program for the average number of signal-setting single-photon counts in the X basis, $\overline{X}_{1,\mu,N}$, such that
\begin{equation}
	\langle{\overline{X}_{1,\mu,N}}\rangle\geq{}q_{\rm X}^{2}p_{\mu}{\mu}^{-}e^{-\mu^{-}}{y'}_{1,\mu,N},
\end{equation}
where the apostrophe here denotes that we refer to the X basis.

On the other hand, the average number of signal-setting single-photon error counts in the X basis is $\overline{E}_{1,\mu,N}=\sum_{k=1}^{N}E_{1,\mu}^{(k)}/N$, with $E_{1,\mu}^{(k)}=E_{\mu}^{(k)}\mathds{1}_{\{n_{k}=1\}}$. Since $\bigl\langle{E_{1,\mu}^{(k)}}\bigr\rangle=q_{\rm X}^{2}p_{\mu}p^{(k)}(1|\mu)H_{1,\mu}^{(k)}\leq{}q_{\rm X}^{2}p_{\mu}{\mu}^{+}e^{-\mu^{+}}H_{1,\mu}^{(k)}$, averaging over $k$ it follows that
\begin{equation}
	\langle{\overline{E}_{1,\mu,N}}\rangle\leq{}q_{\rm X}^{2}p_{\mu}{\mu}^{+}e^{-\mu^{+}}h_{1,\mu,N},
\end{equation}
and an upper bound $h_{1,\mu,N}^{\rm U}$ on $h_{1,\mu,N}$ is reached by the following linear program:
\begin{gather}\label{lp_2}
	\max\quad h_{1,\mu,N}\\
	\begin{aligned}
		\textup{s.t.}\hspace{.3cm}&\frac{\left\langle{\overline{E}_{a,N}}\right\rangle}{q_{\rm X}^{2}p_{a}}\geq  e^{-a^{+}}h_{0,a}+\sum_{n=1}^{n_{\rm cut}}\frac{e^{-a^{-}}a^{-\hspace{.06cm}n}}{n!}h_{n,a,N}\hspace{.2cm}(a\in{}A),\nonumber\\
		&\frac{\left\langle{\overline{E}_{a,N}}\right\rangle}{q_{\rm X}^{2}p_{a}}\leq  1-e^{-a^{+}}+e^{-a^{-}}h_{0,a}-\sum_{n=1}^{n_{\rm cut}}\frac{e^{-a^{+}}a^{+\hspace{.06cm}n}}{n!}\left(1-h_{n,a,N}\right)\hspace{.2cm}(a\in{}A),\nonumber\\
		&t^{+}_{ab,n}+s^{+}_{ab,n}h_{n,a,N}\geq{}{h}_{n,b,N}\hspace{.2cm}(a\in{}A,\ b\in{}A,\ a\neq{}b,\ n=0,\ldots{},n_{\rm cut}),\nonumber\\
		&t^{-}_{ab,n}+s^{-}_{ab,n}h_{n,a,N}\leq{}{h}_{n,b,N}\hspace{.2cm}(a\in{}A,\ b\in{}A,\ a\neq{}b,\ n=0,\ldots{},n_{\rm cut}),\nonumber\\
		&0\leq{}h_{n,a,N}\leq{}1\hspace{.2cm}(a\in{}A,n=0,\ldots,n_{\rm cut}).
	\end{aligned}
\end{gather}

Finally, we remark that, in virtue of the properties of linear optimization~\cite{LP}, $y_{1,\mu,N}^{\rm L}$, ${y'}_{1,\mu,N}^{\rm L}$ and $h_{1,\mu,N}^{\rm U}$ are linear in $\left\langle{\overline{Z}_{a,N}}\right\rangle$, $\left\langle{\overline{X}_{a,N}}\right\rangle$ and $\left\langle{\overline{E}_{a,N}}\right\rangle$, respectively, for all $a\in{}A$, which means that they provide the expectation of certain random variables respectively linear in $\overline{Z}_{a,N}$, $\overline{X}_{a,N}$ and $\overline{E}_{a,N}$. In turn, this implies that the bounds reached by the linear programs can be written as $\langle{\overline{Z}_{1,\mu,N}}\rangle\geq{}\langle{\overline{Z}_{1,\mu,N}^{\rm L}}\rangle$, $\langle{\overline{X}_{1,\mu,N}}\rangle\geq{}\langle{\overline{X}_{1,\mu,N}^{\rm L}}\rangle$ and $\langle{\overline{E}_{1,\mu,N}}\rangle\leq{}\langle{\overline{E}_{1,\mu,N}^{\rm U}}\rangle$, where
\begin{eqnarray}
&&\langle{\overline{Z}_{1,\mu,N}^{\rm L}}\rangle=q_{\rm Z}^{2}p_{\mu}{\mu}^{-}e^{-\mu^{-}}y_{1,\mu,N}^{\rm L},\nonumber \\
&&\langle{\overline{X}_{1,\mu,N}^{\rm L}}\rangle=q_{\rm X}^{2}p_{\mu}{\mu}^{-}e^{-\mu^{-}}{y'}_{1,\mu,N}^{\rm L}\hspace{.4cm}\mathrm{and}\nonumber \\
&&\langle{\overline{E}_{1,\mu,N}^{\rm U}}\rangle=q_{\rm X}^{2}p_{\mu}\mu^{+}e^{-\mu^{+}}h_{1,\mu,N}^{\rm U}
\end{eqnarray}
for all $N$. This feature is crucial to justify the asymptotic approximation of the secret key rate presented next.
\subsection{Asymptotic approximation of the secret key rate}\label{asymptotic_rate}
The linear programs of Sec.~\ref{LP} provide suitable bounds on the expectations of the relevant experimental averages, namely, the average number of signal-setting single-photon counts in the Z (X) basis after $N$ transmission rounds, $\overline{Z}_{1,\mu,N}$ ($\overline{X}_{1,\mu,N}$), and the average number of signal-setting single-photon error counts in the X basis after $N$ transmission rounds, $\overline{E}_{1,\mu,N}$. The bounds are of the form
\begin{equation}\label{expectations} \langle{\overline{Z}_{1,\mu,N}}\rangle\geq{}\langle{\overline{Z}_{1,\mu,N}^{\rm L}}\rangle,\hspace{.2cm} \langle{\overline{X}_{1,\mu,N}}\rangle\geq{}\langle{\overline{X}_{1,\mu,N}^{\rm L}}\rangle\hspace{.2cm}\mathrm{and}\hspace{.2cm}\langle{\overline{E}_{1,\mu,N}}\rangle\leq{}\langle{\overline{E}_{1,\mu,N}^{\rm U}}\rangle
\end{equation}
for all $N$, where $\overline{Z}_{1,\mu,N}^{\rm L}$ ($\overline{X}_{1,\mu,N}^{\rm L}$) is a linear combination of the experimentally observed Z (X) basis gains, $\{\overline{Z}_{a,N}\}_{a\in{}A}$ ($\{\overline{X}_{a,N}\}_{a\in{}A}$), and $\overline{E}_{1,\mu,N}^{\rm U}$ is a linear combination of the experimentally observed numbers of errors in the X basis, $\{\overline{E}_{a,N}\}_{a\in{}A}$.

However, the finite secret key rate relies on statistical bounds of the experimental averages themselves, say
\begin{equation}\label{latter} P\left(\overline{Z}_{1,\mu,N}<\overline{Z}_{1,\mu,N}^{\mathrm{L},\epsilon_{1}}\right)\leq{}\epsilon_{1},\hspace{.2cm} P\left(\overline{X}_{1,\mu,N}<\overline{X}_{1,\mu,N}^{\mathrm{L},\epsilon_{2}}\right)\leq{}\epsilon_{2}\hspace{.2cm}\mathrm{and}\hspace{.2cm}P\left(\overline{E}_{1,\mu,N}>\overline{E}_{1,\mu,N}^{\mathrm{U},\epsilon_{3}}\right)\leq{}\epsilon_{3}
\end{equation}
for given failure probabilities $\epsilon_{1}$, $\epsilon_{2}$ and $\epsilon_{3}$ (see the Methods Sec.~\ref{finite} for a summarized derivation of the finite secret key rate). Crucially, in the absence of intensity correlations or side-channels possibly leaking the intensity setting information, commutativity allows to consider the so-called counterfactual setting, in which case the latter bounds ---Eq.~(\ref{latter})--- are obtained from the former ---Eq.~(\ref{expectations})--- via concentration inequalities for independent random variables, such as Chernoff's~\cite{Chernoff} or Hoeffding's~\cite{Hoeffding}. Precisely, the independence of the relevant indicator variables attached to the detection events is enforced because, in the counterfactual setting, the intensities are randomly selected a posteriori (and thus decoupled) of the detection events. On the contrary, intensity correlations invalidate the counterfactual setting argument, in so invalidating the usage of the above concentration inequalities too. Nevertheless, propositions 1 and 2 in the Methods Sec.~\ref{propositions} establish that, as long as the variance of the experimental averages tends to zero as $N$ tends to infinity, any violation of the equations
\begin{equation}\label{observables} \overline{Z}_{1,\mu,N}\geq{}\overline{Z}_{1,\mu,N}^{\rm L},\hspace{.2cm}\overline{X}_{1,\mu,N}\geq{}\overline{X}_{1,\mu,N}^{\rm L}\hspace{.2cm}\mathrm{and}\hspace{.2cm}\overline{E}_{1,\mu,N}\leq{}\overline{E}_{1,\mu,N}^{\rm U}
\end{equation}
---no matter how small--- has an asymptotically null probability of occurring.

This feature legitimizes the use of the bounds of Eq.~(\ref{observables}) to asymptotically approximate the secret key rate, by plugging them into the finite secret key rate formula ---Eq.~(\ref{keyrate}) in the Methods Sec.~\ref{finite}---. This yields
\begin{eqnarray}\label{signature_trick}
	&&K_{N}\approx\overline{Z}_{1,\mu,N}^{\rm L}\left[1-h\left(\frac{\overline{E}_{1,\mu,N}^{\rm U}}{\overline{X}_{1,\mu,N}^{\rm L}}+\sqrt{\frac{\left(\overline{X}_{1,\mu,N}^{\rm L}+\overline{Z}_{1,\mu,N}^{\rm L}\right)\left(\overline{Z}_{1,\mu,N}^{\rm L}+1/N\right)}{2N{\overline{Z}_{1,\mu,N}^{\rm L\hspace{.05cm}2}}\overline{X}_{1,\mu,N}^{\rm L}}\log\left(\frac{1}{\epsilon_{\rm S}}\right)}\right)\right]\nonumber\\
	&&-f_{\rm EC}\overline{Z}_{\mu,N}h(E_{\rm tol})-\frac{1}{N}\log\left(\frac{1}{\epsilon_{\rm cor}\epsilon^{2}_{\rm PA}\delta}\right),
\end{eqnarray}
%take the finite key formula and plug in the bounds on the expectations but removing the expectations! That's what everyone does, but rigorously
for large enough $N$, with secrecy parameter $\epsilon_{\rm sec}\approx{}2\epsilon_{\rm S}+\epsilon_{\rm PA}+\delta$
%the parameter estimation error can be made as small as desired
and correctness parameter $\epsilon_{\rm cor}$ (see the Methods Sec.~\ref{finite} for the definitions of the security parameters $\epsilon_{\rm S}$, $\epsilon_{\rm PA}$ and $\delta$). Also, $h(\cdot)$ denotes the binary entropy function, $f_{\rm EC}$ is the efficiency of the error correction protocol and $E_{\rm tol}$ is a threshold bit error rate. Note that Eq.~(\ref{signature_trick}) assumes that Alice and Bob use the Z (X) basis events for key generation (parameter estimation).

Having reached this stage, one can remove both the dependence on $N$ and on the security parameters by neglecting the Serfling deviation term (which scales as $N^{-1/2}$) and the finite key term $\log\left(1/\epsilon_{\rm cor}\epsilon^{2}_{\rm PA}\delta\right)/N$, in so obtaining the final asymptotic secret key rate formula
\begin{equation}\label{asymptotic}
	K_{\infty}=\overline{Z}_{1,\mu,N}^{\rm L}\left[1-h\left(\frac{\overline{E}_{1,\mu,N}^{\rm U}}{\overline{X}_{1,\mu,N}^{\rm L}}\right)\right]-f_{\rm EC}\overline{Z}_{\mu,N}h(E_{\rm tol}).
\end{equation}

Notably, as mentioned, the usefulness of this asymptotic approximation is subject to the non-trivial condition that the variance of the experimental averages vanishes as $N$ tends to infinity. Precisely, for a sequence $\{X_{j}\}$ with successive averages $\overline{X}_{N}=\sum_{j=1}^{N}X_{j}/N$, we have that $Var[\overline{X}_{N}]=\sum_{i=1}^{N}Var[X_{i}]/N^{2}+2\sum_{i=1}^{N}\sum_{j>i}^{N}Cov[X_{i},X_{j}]/N^{2}$. Then, since $\lim_{N\to\infty}\sum_{i=1}^{N}Var[X_{i}]/N^{2}\leq{}\lim_{N\to\infty}\max_{i}\{Var[X_{i}]\}/N=0$, it follows that $\lim_{N\to\infty}Var[\overline{X}_{N}]=2\times\lim_{N\to\infty}\sum_{i=1}^{N}\sum_{j>i}^{N}Cov[X_{i},X_{j}]/N^{2}$, as long as the latter is finite. Thus, in particular, Eq.~(\ref{asymptotic}) is an asymptotic approximation of the secret key rate provided that
\begin{equation}\label{condition}
	\lim_{N\to\infty}\sum_{i=1}^{N}\sum_{j>i}^{N}\frac{Cov[X_{i},X_{j}]}{N^{2}}=0
\end{equation}
for the relevant sequences $X_{k}\in\{Z_{a}^{(k)},X_{a}^{(k)},E_{a}^{(k)},{Z}_{1,a}^{(k)},{X}_{1,a}^{(k)},{E}_{1,a}^{(k)}\}_{a\in{}A}$ (\textit{i.e.,} provided that the preconditions of propositions 1 and 2 in the Methods Sec.~\ref{propositions} hold). For instance, if $Cov\left[X_{i},X_{j}\right]=0$ for all $i$, $j$ with $|i-j|>\zeta$ ---where $\zeta$ denotes a finite round difference--- the condition holds despite the fact that the $X_{j}$ may be dependent and non-identically distributed. %Lastly, it is crucial to notice that Eq.~(\ref{condition}) is not only sufficient to assure the asymptotic validity of $K_{\infty}$, but also necessary, as discussed in the Methods Sec.~\ref{propositions}.
%Formally, we did not prove the necessary character!
\subsection{Simulations}\label{simulations}
In the absence of real data, we fix the experimental inputs $\overline{Z}_{a,N}/q_{\rm Z}^{2}p_{a}$, $\overline{X}_{a,N}/q_{\rm X}^{2}p_{a}$ and $\overline{E}_{a,N}/q_{X}^{2}p_{a}$ of the linear programs to their expected values according to a typical channel model. Importantly, although the security analysis contemplates intensity correlations, we adopt a standard channel and transmitter model without correlations for ease of comparison with prior work. In particular, let $\eta_{\rm det}$ ($\eta_{\rm ch}=10^{-\alpha_{\rm att}L/10}$) denote the detection efficiency of Bob's detectors (transmittance of the channel), where $\alpha_{\rm att}$ (dB/km) is the attenuation coefficient of the channel and $L$ (km) is the distance between Alice's and Bob's labs. Also, let $p_{\rm d}$ ($\delta_{\rm A}$) stand for the dark count probability of each of Bob's photo-detectors (polarization misalignment occurring in the channel). The model is~\cite{Lim,Zapatero}
\begin{eqnarray}\label{model}
	&&\frac{\left\langle{\overline{Z}_{a,N}}\right\rangle}{q_{\rm Z}^{2}p_{a}}=\frac{\left\langle{\overline{X}_{a,N}}\right\rangle}{q_{\rm X}^{2}p_{a}}=1-(1-p_{\rm d})^{2}e^{-\eta{}a}\hspace{.2cm}\mathrm{and} \nonumber \\
	&&\frac{\left\langle{\overline{E}_{a,N}}\right\rangle}{q_{X}^{2}p_{a}}=\frac{\left\langle{\overline{E}_{a,N(\rm Z)}}\right\rangle}{q_{Z}^{2}p_{a}}=\frac{p_{\rm d}^{2}}{2}+p_{\rm d}(1-p_{\rm d})\bigl(1+h_{\eta,a,\delta_{\rm A}}\bigr)\nonumber \\
	&&+(1-p_{\rm d})^{2}\left(\frac{1}{2}+h_{\eta,a,\delta_{\rm A}}-\frac{1}{2}e^{-\eta{}a}\right)
\end{eqnarray}
for $a\in{}A$, where $\eta=\eta_{\rm det}\eta_{\rm ch}$ and we define $h_{\eta,a,\delta_{\rm A}}=\bigl(e^{-\eta{}a\cos^2\delta_{\rm A}}-e^{-\eta{}a\sin^2\delta_{\rm A}}\bigr)/2$. Also, we introduce the variable $\overline{E}_{a,N(\rm Z)}$, which is equivalent to $\overline{E}_{a,N}$ but referred to the Z basis error clicks. Note that Eq.~(\ref{model}) accounts for the fact that multiple clicks are randomly assigned to a specific detection outcome. For simplicity, the tolerated bit error rate of the sifted key is set to $E_{\rm tol}=\langle{\overline{E}_{\mu,N(\rm Z)}}\rangle/\langle{\overline{Z}_{\mu,N}}\rangle$.

In addition, we remark that the channel model can also be used to select reference values $\tilde{y}_{n,a,N}$ and $\tilde{h}_{n,a,N}$ for the evaluation of the linearized CS constraints of Sec.~\ref{LP}. Nevertheless, one could also choose the reference values based on previous executions of the protocol instead. Note that, in a real QKD experiment, these reference values would be required for the parameter estimation to go through, and so the secret key rate would be sensitive to the selected $\tilde{y}_{n,a,N}$ and $\tilde{h}_{n,a,N}$. The reader is referred to Appendix~\ref{reference_values} for the explicit formulas of $\tilde{y}_{n,a,N}$ and $\tilde{h}_{n,a,N}$ that we use, matching the typical channel model under consideration. Alternatively, in Appendix~\ref{TD} we provide a looser analysis based on the trace distance argument~\cite{Nielsen}, which does not rely on the selection of any reference values.

Either way, plugging Eq.~(\ref{model}) into Eq.~(\ref{lp_1}) and Eq.~(\ref{lp_2}), we find that the asymptotic secret key rate formula $K_{\infty}$ (presented in Eq.~(\ref{asymptotic})) does not depend either on the probability of the decoy settings, $p_{\nu}$ and $p_{\omega}$, or on the probability of selecting the X basis, $q_{\rm X}$, in such a way that setting $p_{\mu}\approx{}1$ and $q_{\rm Z}\approx{}1$ maximizes $K_{\infty}$ with the typical channel model under consideration. This feature corroborates the intuition that, as $N$ increases, one can devote larger and larger fractions of rounds to key generation without compromising the tightness of the parameter estimation. Lastly, regarding the experimental parameters, we list them below. We take $\eta_{\rm det}=0.65$, $p_{\rm d}=7.2\times{}10^{-8}$ ---both values matching the recent experiment reported in~\cite{Pan1}---, a typical attenuation coefficient $\alpha_{\rm att}=0.2$ dB/km and a standard error correction efficiency of $f_{\rm EC}=1.16$. Regarding the misalignment, we take $\delta_{\rm A}=0.08$ for illustration purposes. Also, we fix the weakest intensity setting to $\omega=10^{-4}$ for the numerics, and we numerically optimize $\mu$ and $\nu$ to maximize $K_{\infty}$ for each value of the distance $L$. Lastly, three different correlation ranges are contemplated, $\xi=1$, $\xi=2$ and $\xi=5$, each of which is combined with various values of the maximum relative deviation $\delta_{\rm max}$ (see assumptions 2 and 3 in Sec.~\ref{assumptions} for the definitions of $\xi$ and $\delta_{\rm max}$).

\begin{figure}[!htbp]
	\centering 
	\includegraphics[width=9.6cm,height=8.1cm]{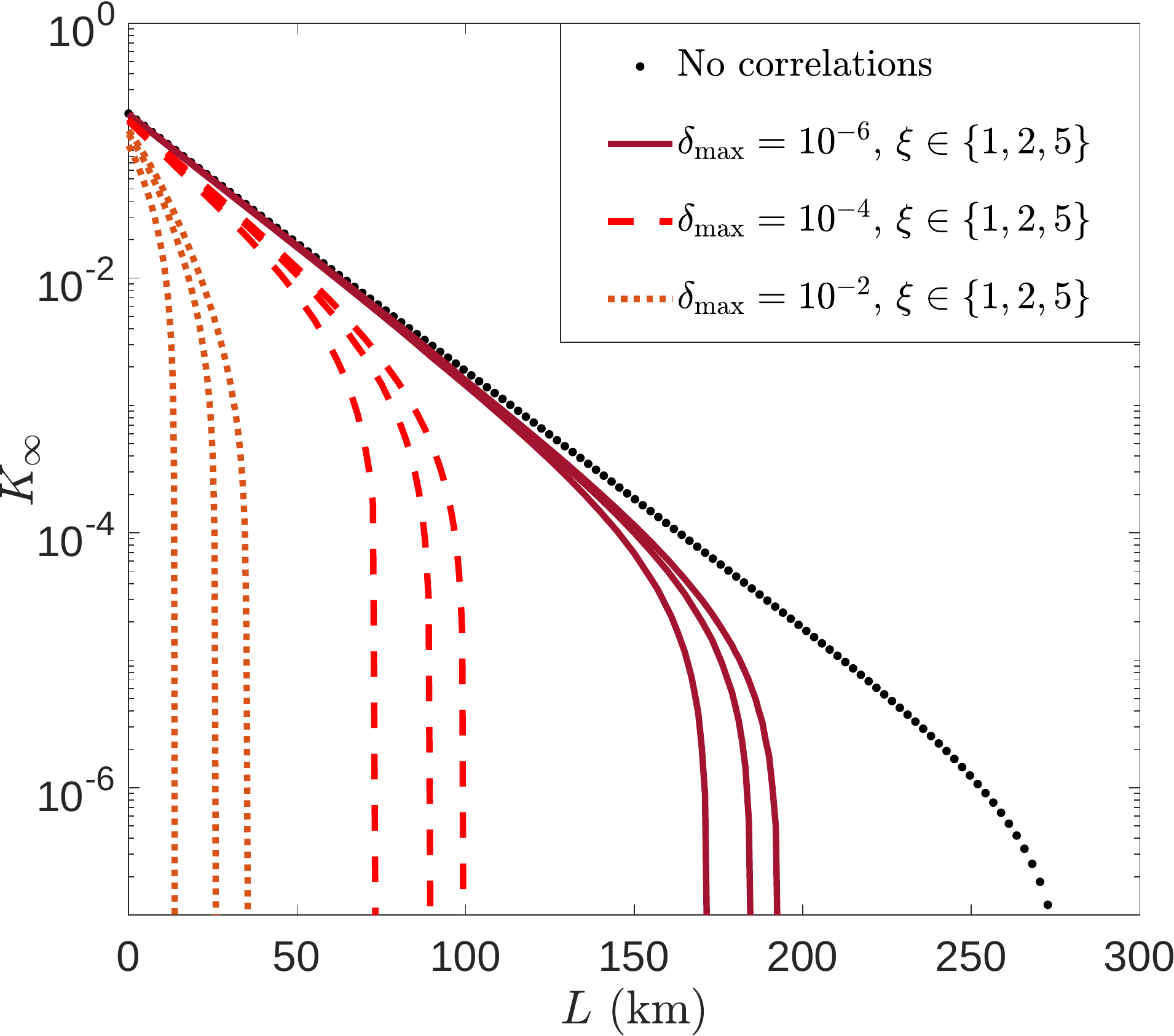}\\
	\caption{QKD performance in the presence of finite range intensity correlations. The figure shows the asymptotic secret key rate, $K_{\infty}$, as a function of the distance, $L$, when performing the parameter estimation with the linearized CS constraint. For illustration purposes, we consider three different values of the maximum relative deviation between intensity settings ($a_{k}$) and actual intensities ($\alpha_{k}$), $\delta_{\rm max}\in\{10^{-6},10^{-4},10^{-2}\}$, and three correlation ranges are contemplated in each case, $\xi\in\{1,2,5\}$. Moreover, for comparison purposes, we show the attainable secret key rate in the absence of intensity correlations as well (dotted black line). The experimental parameters are fixed as specified in the main text.}
	\label{fig:model-independent}
\end{figure}

The rate-distance performance with the above considerations is shown in Fig.~\ref{fig:model-independent}. As seen in the figure, intensity correlations strongly limit the maximum distance attainable for QKD, and the secret key rate is notably sensitive to the deviation parameter, $\delta_{\rm max}$. On the contrary, as long as moderate values are considered for the correlation range $\xi$, the effect of this parameter on the secret key rate is softer. Finally, for completeness, in Appendix~\ref{deterministic} we show that an enhancement of the secret key rate is possible by assuming deterministic intensity correlations, in contrast to the model-independent correlations considered so far.
\section{Discussion}\label{Discussion}
Aiming to enhance the performance of QKD systems, it is crucial to develop ultrafast clock rate QKD devices capable of delivering high secret key rates for widespread applications. However, even for GHz clock rates, QKD transmitters exhibit intensity correlations~\cite{Geneve,Tomita} that invalidate standard decoy-state analyses for parameter estimation. As opposed to many other source imperfections, only limited solutions were known for this security loophole so far~\cite{Tomita,Nagamatsu,Mizutani}. In this work, we solve the problem by quantifying the maximum effect of intensity correlations in the security of QKD. For this purpose, we introduce two experimental-friendly security parameters that allow to characterize arbitrary correlations in the IM. Importantly, our technique builds on a result that we refer to as the Cauchy-Schwarz constraint (recently used in~\cite{Pereira,Navarrete}), which provides tighter bounds on the indistinguishability of non-orthogonal quantum states than the well-known trace distance argument~\cite{Nielsen}.

For illustration purposes, our analysis is dedicated to the standard decoy-state BB84 protocol with one signal and two decoy settings~\cite{Lim}, although we remark that our results can be easily generalized to deal with other variants of the protocol, or even with the decoy-state measurement-device-independent (MDI) QKD scheme~\cite{MDI1}.

As a related contribution, we present a non-standard derivation of the asymptotic limit, in so revealing a necessary condition to justify the ubiquitous asymptotic formula. Crucially, this condition becomes non-trivial in the most general context of coherent attacks and arbitrary pulse correlations. Nevertheless, if, for instance, Eve's attack does not interrelate arbitrarily distant detection events ---but the interaction is limited to a finite round difference--- the condition holds. In this regard, it is not unreasonable to conjecture that, as long as the correlation range of the light pulses is not larger than $\xi$, Eve may reach an optimal cheating strategy by attacking blocks of maximum round difference $\zeta=\xi$. Indeed, a hypothesis of this kind might pave the way for a finite key analysis of the problem, which is the natural direction to follow. In any case, whether or not this conjecture is true, the solution here presented clearly provides an insightful step towards foolproof security of high speed QKD systems, with their imperfections.

\section{Methods}\label{Methods}
\subsection{Cauchy-Schwarz constraint}\label{Cauchy}
The CS constraint is stated as follows~\cite{Pereira,Lo-Preskill}.\\

\textbf{Theorem}~\cite{Pereira}. Let $\ket{u}$ and $\ket{v}$ be pure states of a certain quantum system. Then, for all positive operators $\hat{O}\leq{I}$,
\begin{equation}\label{CS_constraint}
G_{-}\left(\bra{u}\hat{O}\ket{u},\abs{\braket{v}{u}}^{2}\right)\leq{\bra{v}\hat{O}\ket{v}}\leq{}G_{+}\left(\bra{u}\hat{O}\ket{u},\abs{\braket{v}{u}}^{2}\right),
\end{equation}
where the functions $G_{\pm}$ are defined in Eq.~(\ref{Gs}). As pointed out in Sec.~(\ref{main_result}), for all $k=1,\ldots,N$, all $n\in\mathbb{N}$, and any given pair of settings, $a\in{}A$ and $b\in{}A$ with $b\neq{}a$, Eq.~(\ref{CS_constraint}) allows to constrain the maximum deviation that Eve can induce between the $n$-photon yields $p^{(k)}(\mathrm{click}|n,a,\mathrm{Z},\mathrm{Z})=p^{(k)}(\mathrm{click}|n,a,\mathrm{Z})$ and $p^{(k)}(\mathrm{click}|n,b,\mathrm{Z},\mathrm{Z})=p^{(k)}(\mathrm{click}|n,b,\mathrm{Z})$, where we have invoked the basis-independent detection efficiency assumption to remove the conditioning on Bob's basis choice. Here, we derive the specific constraint, namely, Eq.~(\ref{deviation}), which contemplates fully general coherent attacks and finite range intensity correlations.

In an entanglement based view of the protocol, the global input state describing all the protocol rounds reads
\begin{equation}\label{global}
\ket{\Psi}=\left[\sum_{a_{1}^{N}}\sum_{x_{1}^{N}}\sum_{r_{1}^{N}}\biggl(\prod_{i=1}^{N}\sqrt{\frac{p_{a_{i}}q_{x_{i}}}{2}}\biggr)\left(\bigotimes_{i=1}^{N}\ket{a_{i},x_{i}}_{A_{i}}\ket{r_{i}}_{A^{'}_{i}}\ket{\psi^{x_{i},r_{i}}_{\vec{a}_{i}}}_{B_{i}C_{i}}\right)\right]\otimes\ket{0}_{E},
\end{equation}
where we introduce the notation $a_{1}^{N}=a_{1}\ldots{}a_{N}$, and equivalently for $x_{1}^{N}$ and $r_{1}^{N}$. Also, for all $i$, $\bigl\{\ket{a_{i},x_{i}}_{A_{i}}|a_{i}\in{}A,x_{i}\in{}M\bigr\}$ and $\bigl\{\ket{r_{i}}_{{A}^{'}_{i}}|r_{i}\in\mathbb{Z}_{2}\bigr\}$ are orthonormal bases of Alice's $i$-th registers, $A_{i}$ and $A^{'}_{i}$. Similarly, we define
\begin{equation}\label{purified_PRWCP}
\ket{\psi^{x_{i},r_{i}}_{\vec{a}_{i}}}_{B_{i}C_{i}}=\sum_{n_{i}=0}^{\infty}\sqrt{p_{n_{i}}\lvert_{\vec{a}_{i}}}\ket{t_{n_{i}}}_{C_{i}}\ket{n_{i}^{x_{i},r_{i}}}_{B_{i}},
\end{equation}
where $C_{i}$ denotes an inaccessible purifying system with orthonormal basis $\bigl\{\ket{t_{n_{i}}}_{C_{i}}|n_{i}\in{}\mathbb{N}\bigr\}$ for all $i$ ($C_{i}$ stores the photon number information of the $i$-th signal that Alice sends to Bob), $B_{i}$ denotes the system delivered to Bob ($\ket{n_{i}^{x_{i},r_{i}}}_{B_{i}}$ standing for a Fock state with $n_{i}$ photons encoding the BB84 polarization state defined by $(x_{i},r_{i})$), and  the photon number statistics $p_{n_{i}}\lvert_{\vec{a}_{i}}$ are defined in Eq.~(\ref{statistics}). Lastly, $\ket{0}_{E}$ in Eq.~(\ref{global}) stands for the initial state of Eve's ancillary system.

If we denote Eve's coherent interaction with systems $B_{1},\ldots,B_{N}$ and $E$ by $\hat{U}_{BE}$ ---such that $\hat{U}_{BE}\ket{\Psi}$ represents the global state prior to Bob's measurements--- and refer to Bob's ``click'' POVM element in round $k$ as $\hat{M}_{B_{k}}^{\rm click}=\mathds{1}_{B_{k}}-\hat{M}_{B_{k}}^{f}$, the joint probability $p^{(k)}(\mathrm{click},n,a,\mathrm{Z})$ is computed as
\begin{eqnarray}\label{joint}
p^{(k)}(\mathrm{click},n,a,\mathrm{Z})&=&\Tr\left\{\hat{P}_{\ket{a,\mathrm{Z},t_{n}}_{A_{k}C_{k}}}\hat{M}_{B_{k}}^{\rm click}\hspace{.1cm}\hat{U}_{BE}\ketbra{\Psi}{\Psi}\hat{U}^{\dagger}_{BE}\right\}\nonumber \\
&=&\Tr\left\{\hat{U}^{\dagger}_{BE}\hat{M}_{B_{k}}^{\rm click}\hspace{.1cm}\hat{U}_{BE}\hat{P}_{\ket{a,\mathrm{Z},t_{n}}_{A_{k}C_{k}}}\ketbra{\Psi}{\Psi}\hat{P}_{\ket{a,\mathrm{Z},t_{n}}_{A_{k}C_{k}}}\right\} \nonumber \\
&=&\Tr_{\underline{A_{k}C_{k}},{A}_{1}^{'N}B_{1}^{N}E}\left\{\hat{U}^{\dagger}_{BE}\hspace{.1cm}\hat{M}_{B_{k}}^{\rm click}\hspace{.1cm}\hat{U}_{BE}\ketbra{\widetilde{\Psi}^{(k)}_{a,\mathrm{Z},n}}{\widetilde{\Psi}^{(k)}_{a,\mathrm{Z},n}}\right\},
\end{eqnarray}
%Formally, all operators act on the global hilbert space. For this, their definitions are the obvious ones, but tensor-multiplied by the identity acting on the rest of the subsystems. We abuse of this notation.
where $\hat{P}_{\ket*{a,\mathrm{Z},t_{n}}_{A_{k}C_{k}}}=\ketbra{a,\mathrm{Z}}{a,\mathrm{Z}}_{A_{k}}\otimes\ketbra{t_{n}}{t_{n}}_{C_{k}}$, $\underline{A_{k}C_{k}}=\{A_{j}C_{j}|j\neq{}k\}$, and we have introduced the unnormalized pure state
\begin{equation}\label{unnormalized_state}
\ket{\widetilde{\Psi}^{(k)}_{a,\mathrm{Z},n}}=\bra{a,\mathrm{Z}}_{A_{k}}\bra{t_{n}}_{C_{k}}\ket*{\Psi}.
\end{equation}
Note that, in the derivation of Eq.~(\ref{joint}), we make use of the fact that projection operators are ``self-squared'', together with the cyclic property of the trace and straightforward commutation relations. Then, we trace out systems $A_{k}$ and $C_{k}$ explicitly, in order to obtain the necessary input of the CS constraint later on.

Further defining $\ket*{{\Psi}^{(k)}_{a,\mathrm{Z},n}}={\ket*{\widetilde{\Psi}^{(k)}_{a,\mathrm{Z},n}}}/\Vert{\ket*{\widetilde{\Psi}^{(k)}_{a,\mathrm{Z},n}}}\Vert$,
Eq.~(\ref{joint}) can be restated as 
\begin{equation}\label{joint 2}
p^{(k)}(\mathrm{click},n,a,\mathrm{Z})={\left\Vert{\ket{\widetilde{\Psi}^{(k)}_{a,\mathrm{Z},n}}}\right\Vert}^{2}\Tr\left\{\hat{U}^{\dagger}_{BE}\hspace{.1cm}\hat{M}_{B_{k}}^{\rm click}\hspace{.1cm}\hat{U}_{BE}\ketbra{{\Psi}^{(k)}_{a,\mathrm{Z},n}}{{\Psi}^{(k)}_{a,\mathrm{Z},n}}\right\},
\end{equation}
and since $p^{(k)}(n,a,\mathrm{Z})=\Tr\left\{\hat{P}_{\ket{a,\mathrm{Z},t_{n}}_{A_{k}C_{k}}}\hat{U}_{BE}\ketbra{\Psi}{\Psi}\hat{U}^{\dagger}_{BE}\right\}={\left\Vert{\ket{\widetilde{\Psi}^{(k)}_{a,\mathrm{Z},n}}}\right\Vert}^{2}$, it follows from Bayes rule that
\begin{equation}\label{conditional}
p^{(k)}(\mathrm{click}|n,a,\mathrm{Z})=\Tr\left\{\hat{O}^{(k)}_{\rm click}\ketbra{{\Psi}^{(k)}_{a,\mathrm{Z},n}}{{\Psi}^{(k)}_{a,\mathrm{Z},n}}\right\}=\bra{{\Psi}^{(k)}_{a,\mathrm{Z},n}}\hat{O}^{(k)}_{\rm click}\ket{{\Psi}^{(k)}_{a,\mathrm{Z},n}},
\end{equation}
where $\hat{O}^{(k)}_{\rm click}=\hat{U}^{\dagger}_{BE}\hspace{.1cm}\hat{M}_{B_{k}}^{\rm click}\hspace{.1cm}\hat{U}_{BE}$. Recalling that $p^{(k)}(\mathrm{click}|n,a,\mathrm{Z})=p^{(k)}(\mathrm{click}|n,a,\mathrm{Z},\mathrm{Z})=:Y_{n,a}^{(k)}$, which is the $n$-photon yield of round $k$ associated to the intensity setting $a$, it follows from Eq.~(\ref{CS_constraint}) that
\begin{equation}\label{deviation_copy}
G_{-}\left(Y_{n,a}^{(k)},\abs{\braket{{\Psi}^{(k)}_{b,\mathrm{Z},n}}{{\Psi}^{(k)}_{a,\mathrm{Z},n}}}^{2}\right)\leq{}Y_{n,b}^{(k)}\leq{}G_{+}\left(Y_{n,a}^{(k)},\abs{\braket{{\Psi}^{(k)}_{b,\mathrm{Z},n}}{{\Psi}^{(k)}_{a,\mathrm{Z},n}}}^{2}\right)
\end{equation}
for all $n\in{}\mathbb{N}$, $a\in{}A$, $b\in{}A$ ($b\neq{}a$) and $k=1,\ldots,N$, and the bounds are tighter the closer $\abs*{\braket*{{\Psi}^{(k)}_{b,\mathrm{Z},n}}{{\Psi}^{(k)}_{a,\mathrm{Z},n}}}^{2}$ is to 1. We recall that the interpretation is simple: even if Eve fine-tunes her global unitary $\hat{U}_{BE}$ focusing only in round $k$, aiming to maximally deviate $Y_{n,a}^{(k)}$ and $Y_{n,b}^{(k)}$, such a deviation is subject to Eq.~(\ref{deviation_copy}).

In short, evaluating Eq.~(\ref{deviation_copy}) requires to lower bound $\abs*{\braket*{{\Psi}^{(k)}_{b,\mathrm{Z},n}}{{\Psi}^{(k)}_{a,\mathrm{Z},n}}}^{2}$, which we do next. For $k=2,\ldots,N-1$ (the cases $k=1$ and $k=N$ will be discussed separately), we have that
\begin{eqnarray}\label{unnormalized}
&&\ket{\widetilde{\Psi}^{(k)}_{a,\mathrm{Z},n}}=\sqrt{\frac{q_{\rm Z}p_{a}}{2^{N}}}\left[\sum_{\underline{a_{k}}}\sum_{\underline{x_{k}}}\sum_{r_{1}^{N}}\biggl(\prod_{i\neq{k}}\sqrt{p_{a_{i}}q_{x_{i}}}\biggr)\left(\bigotimes_{i=1}^{k-1}\ket{a_{i},x_{i}}_{A_{i}}\ket{r_{i}}_{A^{'}_{i}}\ket{\psi^{x_{i},r_{i}}_{\vec{a}_{i}}}_{B_{i}C_{i}}\right)\times\right.\nonumber\\
&&\left.\left(\sqrt{p_{n}\lvert_{a,\vec{a}_{k-1}}}\ket{r_{k}}_{A^{'}_{k}}\ket{n^{\mathrm{Z},r_{k}}}_{B_{k}}\right)\left(\bigotimes_{i=k+1}^{N}\ket{a_{i},x_{i}}_{A_{i}}\ket{r_{i}}_{A^{'}_{i}}\ket{\psi^{x_{i},r_{i}}_{\vec{a}_{i}(a_{k}=a)}}_{B_{i}C_{i}}\right)\right]\otimes\ket{0}_{E}\nonumber \\
\end{eqnarray}
with $\underline{a_{k}}=\{a_{j}|j\neq{}k\}$ and $\underline{x_{k}}=\{x_{j}|j\neq{}k\}$. Thus, it follows that
\begin{eqnarray}\label{overlap}
&&\braket{\widetilde{\Psi}^{(k)}_{b,\mathrm{Z},n}}{\widetilde{\Psi}^{(k)}_{a,\mathrm{Z},n}}=\frac{q_{\mathrm{Z}}\sqrt{p_{a}p_{b}}}{2^{N}}\sum_{\underline{a_{k}}}\sum_{\underline{x_{k}}}\sum_{r_{1}^{N}}\biggl(\prod_{i\neq{}k}p_{a_{i}}q_{x_{i}}\biggr)\biggl(\sqrt{p_{n}\lvert_{a,\vec{a}_{k-1}}p_{n}\lvert_{b,\vec{a}_{k-1}}}\biggr)\times\nonumber\\
&&\left(\prod_{i=k+1}^{N}{}\braket{\psi^{x_{i},r_{i}}_{\vec{a}_{i}(a_{k}=b)}}{\psi^{x_{i},r_{i}}_{\vec{a}_{i}(a_{k}=a)}}_{B_{i}C_{i}}\right)=q_{\rm Z}\sqrt{p_{a}p_{b}}\sum_{a_{1}^{k-1}}\left(\prod_{i=1}^{k-1}p_{a_{i}}\right)\sqrt{p_{n}\lvert_{a,\vec{a}_{k-1}}p_{n}\lvert_{b,\vec{a}_{k-1}}}\times\nonumber\\
&&\left[\sum_{a_{k+1}^{N}}\biggl(\prod_{i=k+1}^{N}p_{a_{i}}\braket{\psi_{\vec{a}_{i}(a_{k}=b)}}{\psi_{\vec{a}_{i}(a_{k}=a)}}_{B_{i}C_{i}}\biggr)\right]
\end{eqnarray}
for $k=2,\ldots,N-1$, where we have made use of the fact that $\braket*{\psi^{x_{i},r_{i}}_{\vec{a}_{i}(a_{k}=b)}}{\psi^{x_{i},r_{i}}_{\vec{a}_{i}(a_{k}=a)}}_{B_{i}C_{i}}$ is independent of $x_{i}$ and $r_{i}$ for all $i$ ---which is straightforward to show from Eq.~(\ref{purified_PRWCP})--- in order to carry out the sums over $\underline{x_{k}}$ and $r_{1}^{N}$: $\sum_{\underline{x_{k}}}\sum_{r_{1}^{N}}\left(\prod_{i\neq{}k}q_{x_{i}}\right)=\sum_{r_{1}^{N}}\left\{\sum_{\underline{x_{k}}}\left(\prod_{i\neq{}k}q_{x_{i}}\right)\right\}=2^{N}$. Also for this reason, in the last equality we have renamed $\braket*{\psi^{x_{i},r_{i}}_{\vec{a}_{i}(a_{k}=b)}}{\psi^{x_{i},r_{i}}_{\vec{a}_{i}(a_{k}=a)}}_{B_{i}C_{i}}$ simply as $\braket*{\psi_{\vec{a}_{i}(a_{k}=b)}}{\psi_{\vec{a}_{i}(a_{k}=a)}}_{B_{i}C_{i}}$. As expected, particularizing $a=b$ in Eq.~(\ref{overlap}), we obtain $\Vert{\ket*{\widetilde{\Psi}^{(k)}_{a,\mathrm{Z},n}}}\Vert^{2}=q_{\rm Z}p_{a}\sum_{a_{1}^{k-1}}\bigl(\prod_{i=1}^{k-1}p_{a_{i}}\bigr)p_{n}\lvert_{a,\vec{a}_{k-1}}$. Thus, for the normalized states, we have
\begin{eqnarray}\label{overlap_2}
&&\braket{{\Psi}^{(k)}_{b,\mathrm{Z},n}}{{\Psi}^{(k)}_{a,\mathrm{Z},n}}=\sum_{a_{1}^{k-1}}\sqrt{p^{(k)}(\vec{a}_{k-1}|n,a,\mathrm{Z})p^{(k)}(\vec{a}_{k-1}|n,b,\mathrm{Z})}\times\nonumber \\
&&\left[\sum_{a_{k+1}^{N}}\biggl(\prod_{i=k+1}^{N}p_{a_{i}}\braket{\psi_{\vec{a}_{i}(a_{k}=b)}}{\psi_{\vec{a}_{i}(a_{k}=a)}}_{B_{i}C_{i}}\biggr)\right]
\end{eqnarray}
for $k=2,\ldots,N-1$, where we have introduced the obvious definition
\begin{equation}\label{amplitude}
p^{(k)}(\vec{a}_{k-1}|n,a,\mathrm{Z})=\frac{\left(\prod_{i=1}^{k-1}p_{a_{i}}\right)p_{n}\lvert_{a,\vec{a}_{k-1}}}{\sum_{a_{1}^{k-1}}\left(\prod_{i=1}^{k-1}p_{a_{i}}\right)p_{n}\lvert_{a,\vec{a}_{k-1}}}.
\end{equation}
Lastly, regarding the extreme rounds (which are excluded from Eq.~(\ref{overlap_2})), explicit calculation shows that
\begin{eqnarray}\label{extreme}
&&\braket{{\Psi}^{(1)}_{b,\mathrm{Z},n}}{{\Psi}^{(1)}_{a,\mathrm{Z},n}}=\sum_{a_{2}^{N}}\left(\prod_{i=2}^{N}p_{a_{i}}\braket{\psi_{\vec{a}_{i}(a_{1}=b)}}{\psi_{\vec{a}_{i}(a_{1}=a)}}_{B_{i}C_{i}}\right),\nonumber \\
&&\braket{{\Psi}^{(N)}_{b,\mathrm{Z},n}}{{\Psi}^{(N)}_{a,\mathrm{Z},n}}=\sum_{a_{1}^{N-1}}\sqrt{p^{(N)}(\vec{a}_{N-1}|n,a,\mathrm{Z})p^{(N)}(\vec{a}_{N-1}|n,b,\mathrm{Z})}.
\end{eqnarray}
We remark that, so far, we have not imposed Assumption 3 on the intensity correlations yet (see Sec.~\ref{assumptions}).
%Assumptions 1 and 2 have been imposed at the moment of defining the general state, although the resulting structure of the photon-number statistics has not been exploited yet.
At this stage, we invoke it by considering a finite correlation range $\xi$, which allows to rewrite Eq.~(\ref{overlap_2}) as
\begin{eqnarray}\label{overlap_finite_range}
&&\braket{{\Psi}^{(k)}_{b,\mathrm{Z},n}}{{\Psi}^{(k)}_{a,\mathrm{Z},n}}=\nonumber\\
&&\sum_{a_{\max\{k-\xi,1\}}^{k-1}}\sqrt{p^{(k)}({a}_{k-1},\ldots,{a}_{\max\{k-\xi,1\}}|n,a,\mathrm{Z})p^{(k)}({a}_{k-1},\ldots,{a}_{\max\{k-\xi,1\}}|n,b,\mathrm{Z})}\nonumber \\
&&\times\left[\sum_{a_{k+1}^{\min\{k+\xi,N\}}}\left(\prod_{i=k+1}^{\min\{k+\xi,N\}}p_{a_{i}}\braket{\psi_{\vec{a}_{i}(a_{k}=b)}}{\psi_{\vec{a}_{i}(a_{k}=a)}}_{B_{i}C_{i}}\right)\right]
\end{eqnarray}
for $k=2,\ldots,N-1$, and similarly for Eq.~(\ref{extreme}).
%\begin{eqnarray}\label{extreme_2}
%&&\braket{{\Psi}^{(1)}_{b,\mathrm{Z},n}}{{\Psi}^{(1)}_{a,\mathrm{Z},n}}=\sum_{a_{2}^{\xi+1}}\left(\prod_{i=2}^{\xi+1}p_{a_{i}}\braket{\psi_{a_{i},\ldots{},b}}{\psi_{a_{i},\ldots,a}}_{B_{i}C_{i}}\right)\hspace{.2cm}\mathrm{and}\nonumber \\
%&&\braket{{\Psi}^{(N)}_{b,\mathrm{Z},n}}{{\Psi}^{(N)}_{a,\mathrm{Z},n}}=\sum_{a_{N-\xi}^{N-1}}\sqrt{p^{(N)}(a_{N-1},\ldots,a_{N-\xi}|n,a,\mathrm{Z})p^{(N)}(a_{N-1},\ldots,a_{N-\xi}|n,b,\mathrm{Z})}.
%\end{eqnarray}
Crucially, we have made use of the fact that $\braket*{\psi_{\vec{a}_{i}(a_{k}=b)}}{\psi_{\vec{a}_{i}(a_{k}=a)}}_{B_{i}C_{i}}=1$ for all $i>k+\xi$ , which is a straightforward consequence of Assumption 3.

Aiming to evaluate Eq.~(\ref{deviation_copy}), one must lower-bound the right-hand side of Eq.~(\ref{overlap_finite_range}). For this purpose, we 
are going to exploit the structure of Eq.~(\ref{statistics}) in order to derive model-independent bounds valid for all correlations models $g_{\vec{a}_{k}}(\alpha_{k})$. Let us address the bracket in Eq.~(\ref{overlap_finite_range}) first. Noticing that $e^{-x}x^{n}$ is strictly decreasing (increasing) for $n=0$ ($n=1,2,\ldots$) in the interval $x\in(0,1)$, from Eq.~(\ref{statistics}) we have that
$p_{0}\lvert_{\vec{a}_{i}(a_{k}=a)}\geq{}e^{-a_{i}^{+}}$ and $p_{n\geq{}1}\lvert_{\vec{a}_{i}(a_{k}=a)}\geq{}e^{-a_{i}^{-}}{a_{i}^{-\ n}}/n!$ for all $a$, such that explicit calculation yields $\braket*{\psi_{\vec{a}_{i}(a_{k}=b)}}{\psi_{\vec{a}_{i}(a_{k}=a)}}_{B_{i}C_{i}}\geq{}1-(e^{-a_{i}^{-}}-e^{-a_{i}^{+}})$. Therefore, it is easy to show that
\begin{equation}\label{bracket}
\sum_{a_{k+1}^{\min\{k+\xi,N\}}}\left(\prod_{i=k+1}^{\min\{k+\xi,N\}}p_{a_{i}}\braket{\psi_{\vec{a}_{i}(a_{k}=b)}}{\psi_{\vec{a}_{i}(a_{k}=a)}}_{B_{i}C_{i}}\right)\geq{}\left[1-\sum_{c\in{}A}p_{c}\left(e^{-c^{-}}-e^{-c^{+}}\right)\right]^{\xi},
\end{equation}
which becomes a global prefactor in Eq.~(\ref{overlap_finite_range}), as it does not depend on the remaining summation indexes $a_{\max\{k-\xi,1\}}$ to $a_{k-1}$. If we now focus on the first row of Eq.~(\ref{overlap_finite_range}), the same monotonicity argument yields
\begin{equation}\label{un-referenced_2}
p^{(k)}({a}_{k-1},\ldots,{a}_{\max\{k-\xi,1\}}|n,a,\mathrm{Z})\geq{}\begin{cases}
p_{a_{k-1}}\ldots{}p_{a_{\max\{k-\xi,1\}}}e^{a^{-}-a^{+}} & \mathrm{if}\hspace{.3cm}n=0\\
p_{a_{k-1}}\ldots{}p_{a_{\max\{k-\xi,1\}}}e^{a^{+}-a^{-}}(a^{-}/a^{+})^{n} & \mathrm{if}\hspace{.3cm}n\geq{}1,
\end{cases}
\end{equation}
such that
\begin{eqnarray}\label{first_row}
&&\sum_{a_{\max\{k-\xi,1\}}^{k-1}}\sqrt{p^{(k)}({a}_{k-1},\ldots,{a}_{\max\{k-\xi,1\}}|n,a,\mathrm{Z})p^{(k)}({a}_{k-1},\ldots,{a}_{\max\{k-\xi,1\}}|n,b,\mathrm{Z})}\geq{}\nonumber \\
&&\begin{cases}
\exp\left\{\frac{a^{-}+b^{-}-(a^{+}+b^{+})}{2}\right\} & \mathrm{if}\hspace{.3cm}n=0\\
\exp\left\{\frac{a^{+}+b^{+}-(a^{-}+b^{-})}{2}\right\}\left(\frac{a^{-}b^{-}}{a^{+}b^{+}}\right)^{n/2} & \mathrm{if}\hspace{.3cm}n\geq{}1  
\end{cases}
\end{eqnarray}
for $k=2,\ldots,N-1$. Now, putting together Eq.~(\ref{bracket}) and Eq.~(\ref{first_row}), we conclude that $\abs*{\braket*{{\Psi}^{(k)}_{b,\mathrm{Z},n}}{{\Psi}^{(k)}_{a,\mathrm{Z},n}}}^{2}\geq{}\tau_{ab,n}^{\xi}$ for all $k=2,\ldots{}N-1$, $n\in{}\mathbb{N}$, $a\in{}A$, $b\in{}A$ and $b\neq{}a$, where $\tau_{ab,n}^{\xi}$ is given in Eq.~(\ref{tau}) of Sec.~(\ref{main_result}). Indeed, in virtue of Eq.~(\ref{extreme}), it is clear that the resulting bound also applies to the extreme rounds $k=1$ and $k=N$ (the bound is simply less tight in these cases).\\

Lastly, to conclude the proof of Eq.~(\ref{deviation}), we need to establish the same result for the $n$-photon error click probabilities too, \textit{i.e.,} we need to show that
\begin{equation}\label{CS_errors}
G_{-}\left(H_{n,a,r}^{(k)},\tau_{ab,n}^{\xi}\right)\leq{}H_{n,b,r}^{(k)}\leq{}G_{+}\left(H_{n,a,r}^{(k)},\tau_{ab,n}^{\xi}\right)
\end{equation}
for all $k=1,\ldots{}N$, $n\in{}\mathbb{N}$, $a\in{}A$, $b\in{}A$ and $b\neq{}a$, where we recall that $H_{n,a,r}^{(k)}=p^{(k)}(\mathrm{err}|n,a,\mathrm{X,X},r)$. For this purpose, note that, following identical steps as those leading to Eq.~(\ref{conditional}), one finds
\begin{equation}\label{conditional 2}
H_{n,a,r}^{(k)}=\Tr\left\{\hat{O}^{(k)}_{\mathrm{X,err},r}\ketbra{{\Psi}^{(k)}_{a,\mathrm{X},r,n}}{{\Psi}^{(k)}_{a,\mathrm{X},r,n}}\right\}=\bra{{\Psi}^{(k)}_{a,\mathrm{X},r,n}}\hat{O}^{(k)}_{\mathrm{X,err},r}\ket{{\Psi}^{(k)}_{a,\mathrm{X},r,n}}
\end{equation}
for
\begin{equation}
\hat{O}^{(k)}_{\mathrm{X,err},r}=\hat{U}^{\dagger}_{BE}\hspace{.1cm}\hat{M}_{B_{k}}^{\mathrm{X},1-r}\hspace{.1cm}\hat{U}_{BE}\hspace{.2cm}\textrm{and}\hspace{.2cm}\ket{{\Psi}^{(k)}_{a,\mathrm{X},r,n}}=\frac{\ket{\widetilde{\Psi}^{(k)}_{a,\mathrm{X},r,n}}}{\left\Vert{\ket{\widetilde{\Psi}^{(k)}_{a,\mathrm{X},r,n}}}\right\Vert},
\end{equation}
where $\ket*{\widetilde{\Psi}^{(k)}_{a,\mathrm{X},r,n}}=\bra{a,\mathrm{X}}_{A_{k}}\bra{r}_{{A}^{'}_{k}}\bra{t_{n}}_{C_{k}}\ket*{\Psi}$. Thus, in virtue of the CS constraint ---given in Eq.~(\ref{CS_constraint})--- we have that
\begin{equation}\label{Kato_errors}
G_{-}\left(H_{n,a,r}^{(k)},\abs{\braket{{\Psi}^{(k)}_{b,\mathrm{X},r,n}}{{\Psi}^{(k)}_{a,\mathrm{X},r,n}}}^{2}\right)\leq{}H_{n,b,r}^{(k)}\leq{}G_{+}\left(H_{n,a,r}^{(k)},\abs{\braket{{\Psi}^{(k)}_{b,\mathrm{X},r,n}}{{\Psi}^{(k)}_{a,\mathrm{X},r,n}}}^{2}\right),
\end{equation}
and Eq.~(\ref{CS_errors}) follows from the fact that $\abs*{\braket*{{\Psi}^{(k)}_{b,\mathrm{X},r,n}}{{\Psi}^{(k)}_{a,\mathrm{X},r,n}}}=\abs*{\braket*{{\Psi}^{(k)}_{b,\mathrm{Z},n}}{{\Psi}^{(k)}_{a,\mathrm{Z},n}}}$ for both $r=0$ and $r=1$ and for any given $n$, $k$, $a$ and $b$, which is easy to show following the same steps that lead to Eq.~(\ref{overlap_2}). 

\subsection{Phase error rate and secret key length in the finite key regime}\label{finite}
The phase error rate is defined as $\phi_{1,\mathrm{Z},N}:=E_{1,\mu,N}^{\mathrm{Z,ph}}/Z_{1,\mu,N}$, where $E_{1,\mu,N}^{\mathrm{Z,ph}}$ is the number of phase errors among all $Z_{1,\mu,N}$ single-photon events contributing to the sifted key. In this regard, we recall that a phase error is a bit error in a virtual entanglement-based protocol where, for the sifted key rounds, the parties measure their ancillas in the X basis instead. Let us define the set of rounds
\begin{equation}
\mathcal{M}_{1,\mu,N}=\mathcal{Z}_{1,\mu,N}\cup\mathcal{X}_{1,\mu,N},\hspace{.2cm}\mathrm{where}\hspace{.2cm}\mathcal{Z}_{1,\mu,N}=\left\{k\middle|Z_{1,\mu}^{(k)}=1\right\}\hspace{.2cm}\mathrm{and}\hspace{.2cm}\mathcal{X}_{1,\mu,N}=\left\{k\middle|X_{1,\mu}^{(k)}=1\right\}.
\end{equation}
The partition of $\mathcal{M}_{1,\mu,N}$ into $\mathcal{Z}_{1,\mu,N}$ and $\mathcal{X}_{1,\mu,N}$ is common to both the actual and the virtual protocol. In the virtual protocol, a specific number of bit errors occurs in $\mathcal{M}_{1,\mu,N}$, given by the number $E_{1,\mu,N}^{\mathrm{Z,ph}}$ of errors in $\mathcal{Z}_{1,\mu,N}$ plus the number $E_{1,\mu,N}$ of errors in $\mathcal{X}_{1,\mu,N}$. Basis-independence of the single-photon states delivered by Alice in the rounds indexed by $\mathcal{M}_{1,\mu,N}$ implies that Eve cannot distinguish test single-photons ($i\in\mathcal{X}_{1,\mu,N}$) from key single-photons ($i\in\mathcal{Z}_{1,\mu,N}$). Moreover, since, in the virtual protocol, Alice and Bob measure their ancillas in the X basis in both types of rounds, for any given round in $\mathcal{M}_{1,\mu,N}$ the probability that it yields an error is independent of its round-type in the virtual protocol. Thus, one can imagine that Eve is inducing the bit errors in $\mathcal{M}_{1,\mu,N}$ first, and later on Alice and Bob randomly select a partition $\mathcal{M}_{1,\mu,N}=\mathcal{Z}_{1,\mu,N}\cup\mathcal{X}_{1,\mu,N}$, such that
\begin{equation}
\left\langle{\phi_{1,\mathrm{Z},N}}\right\rangle=\left\langle{\frac{E_{1,\mu,N}}{X_{1,\mu,N}}}\right\rangle
\end{equation}
and one can derive a statistical upper bound on the difference $|\phi_{1,\mathrm{Z},N}-E_{1,\mu,N}/X_{1,\mu,N}|$ via Serfling's inequality~\cite{Serfling}. For our purposes, the relevant one-sided bound can be stated as~\cite{Tomamichel}
\begin{equation}\label{Serfling}
P\left\{\phi_{1,\mathrm{Z},N}>\frac{\overline{E}_{1,\mu,N}}{\overline{X}_{1,\mu,N}}+\gamma_{\epsilon_{\rm S}}\right\}\leq{}\epsilon_{\rm S}\hspace{.1cm}\mathrm{for}\hspace{.1cm}\gamma_{\epsilon_{\rm S}}=\sqrt{\frac{\left(\overline{X}_{1,\mu,N}+\overline{Z}_{1,\mu,N}\right)\left(\overline{Z}_{1,\mu,N}+1/N\right)}{2N\overline{Z}_{1,\mu,N}^{2}\overline{X}_{1,\mu,N}}\log\left(\frac{1}{\epsilon_{\rm S}}\right)}.
\end{equation}
Now, let the following inequalities be given:
\begin{equation}\label{bounds}
P(\overline{Z}_{1,\mu,N}<\overline{Z}_{1,\mu,N}^{\mathrm{L},\epsilon_{1}})\leq{}\epsilon_{1},\hspace{.2cm}P(\overline{X}_{1,\mu,N}<\overline{X}_{1,\mu,N}^{\mathrm{L},\epsilon_{2}})\leq{}\epsilon_{2},\hspace{.2cm}\mathrm{and}\hspace{.2cm}P(\overline{E}_{1,\mu,N}>\overline{E}_{1,\mu,N}^{\mathrm{U},\epsilon_{3}})\leq{}\epsilon_{3},
\end{equation}
for certain $\overline{Z}_{1,\mu,N}^{\mathrm{L},\epsilon_{1}}$, $\overline{X}_{1,\mu,N}^{\mathrm{L},\epsilon_{2}}$ and $\overline{E}_{1,\mu,N}^{\mathrm{U},\epsilon_{3}}$ that depend on the observables. In virtue of the union bound, it follows from Eq.~(\ref{Serfling}) and Eq.~(\ref{bounds}) that
\begin{equation}\label{Serfling_2}
P\left\{\phi_{1,\mathrm{Z},N}>\frac{E_{1,\mu,N}^{\mathrm{U},\epsilon_{3}}}{X_{1,\mu,N}^{\mathrm{L},\epsilon_{2}}}+\gamma_{\epsilon_{\rm S},\epsilon_{1},\epsilon_{2}}\right\}\leq{}\epsilon_{\rm S}+\sum_{i=1}^{3}\epsilon_{i}
\end{equation}
for
\begin{equation}\label{un-referenced_3}
\gamma_{\epsilon_{\rm S},\epsilon_{1},\epsilon_{2}}=\sqrt{\frac{\left(\overline{X}_{1,\mu,N}^{\mathrm{L},\epsilon_{2}}+\overline{Z}_{1,\mu,N}^{\mathrm{L},\epsilon_{1}}\right)\left(\overline{Z}_{1,\mu,N}^{\mathrm{L},\epsilon_{1}}+1/N\right)}{2N{\overline{Z}_{1,\mu,N}^{\mathrm{L},\epsilon_{1}\hspace{.05cm}2}}\overline{X}_{1,\mu,N}^{\mathrm{L},\epsilon_{2}}}\log\left(\frac{1}{\epsilon_{\rm S}}\right)}.
\end{equation}
%This is the exact same bound of my second project

At this stage, one can present the secret key rate of the decoy-state BB84 protocol under consideration, which relies on a lower bound on $\overline{Z}_{1,\mu,N}$ (presumed in Eq.~(\ref{bounds})) and an upper bound on $\phi_{1,\mathrm{Z},N}$ (Eq.~(\ref{Serfling_2})). Precisely, for any $\delta\in(0,1)$, it is known that privacy amplification allows to extract an $\epsilon_{\rm sec}$-secret, $\epsilon_{\rm cor}$-correct secret key of length~\cite{Lim}
\begin{equation}\label{key_length}
l=Z_{1,\mu,N}^{\mathrm{L},\epsilon_{1}}\left[1-h\left(\frac{E_{1,\mu,N}^{\mathrm{U},\epsilon_{3}}}{X_{1,\mu,N}^{\mathrm{L},\epsilon_{2}}}+\gamma_{\epsilon_{\rm S},\epsilon_{1},\epsilon_{2}}\right)\right]-f_{\rm EC}Z_{\mu,N}h(E_{\rm tol})-\log\left(\frac{1}{\epsilon_{\rm cor}\epsilon^{2}_{\rm PA}\delta}\right)
\end{equation}
as long as $\epsilon_{\rm sec}\geq{}2\varepsilon+\epsilon_{\rm PA}+\delta$, where $f_{\rm EC}$ is the efficiency of the error correction protocol, $Z_{\mu,N}$ provides the length of the sifted key (see Sec.~\ref{decoy-state} for the definition of $Z_{\mu,N}$), $h(\cdot)$ denotes the binary entropy function, $E_{\rm tol}$ is a threshold bit error rate for the error correction, $\epsilon_{\rm PA}$ is the error probability of the privacy amplification and $\varepsilon=\epsilon_{\rm S}+\sum_{i=1}^{3}\epsilon_{i}$ is the parameter estimation error, \textit{i.e.}, an upper bound on the total error probability of the parameter estimation. Of course, the secret key rate is defined as $K_{N}=l/N$. That is to say,
\begin{equation}\label{keyrate}
K_{N}=\overline{Z}_{1,\mu,N}^{\mathrm{L},\epsilon_{1}}\left[1-h\left(\frac{\overline{E}_{1,\mu,N}^{\mathrm{U},\epsilon_{3}}}{\overline{X}_{1,\mu,N}^{\mathrm{L},\epsilon_{2}}}+\gamma_{\epsilon_{\rm S},\epsilon_{1},\epsilon_{2}}\right)\right]-f_{\rm EC}\overline{Z}_{\mu,N}h(E_{\rm tol})-\frac{1}{N}\log\left(\frac{1}{\epsilon_{\rm cor}\epsilon^{2}_{\rm PA}\delta}\right),
\end{equation}
where we have introduced the notation $\overline{Y}=Y/N$.
\subsection{Technical claims on the asymptotic regime}\label{propositions}
The asymptotic secret key rate formula given in Sec.~\ref{asymptotic_rate} builds on the assertion that, as long as the variance of the experimental averages tends to zero as $N\to\infty$, the probability of any finite violation of Eq.~(\ref{observables}) vanishes for $N\to\infty$ too. Propositions 1 and 2 below formally demonstrate this claim.\\

\textbf{Proposition 1.} Let us assume that $\lim_{N\to\infty}Var\left[\overline{Z}_{a,N}\right]=0$ for all $a\in{}A$ and $\lim_{N\to\infty}Var\left[\overline{Z}_{1,\mu,N}\right]=0$. Then, $\lim_{N\to\infty}P(\overline{Z}_{1,\mu,N}\leq{}\overline{Z}_{1,\mu,N}^{\rm L}-\delta)=0$ for all $\delta>0$. The proposition holds too if one replaces Z by X everywhere.\\

Proof. Let us consider the event $E_{\delta,N}=\left\{\overline{Z}_{1,\mu,N}^{\rm L}-\overline{Z}_{1,\mu,N}\geq{}\delta\right\}$. We have
\begin{eqnarray}
E_{\delta,N}&=&\left\{\overline{Z}_{1,\mu,N}^{\rm L}-\bigl\langle{\overline{Z}_{1,\mu,N}^{\rm L}}\bigr\rangle+\left\langle{\overline{Z}_{1,\mu,N}}\right\rangle-\overline{Z}_{1,\mu,N}+\bigl\langle{\overline{Z}_{1,\mu,N}^{\rm L}}\bigr\rangle-\left\langle{\overline{Z}_{1,\mu,N}}\right\rangle\geq{}\delta\right\}\nonumber\\
&\subseteq{}&\left\{\bigl|\overline{Z}_{1,\mu,N}^{\rm L}-\bigl\langle{\overline{Z}_{1,\mu,N}^{\rm L}}\bigr\rangle\bigr|+\abs{\left\langle{\overline{Z}_{1,\mu,N}}\right\rangle-\overline{Z}_{1,\mu,N}}+\bigl\langle{\overline{Z}_{1,\mu,N}^{\rm L}}\bigr\rangle-\left\langle{\overline{Z}_{1,\mu,N}}\right\rangle\geq{}\delta\right\}\nonumber\\
&\subseteq{}&\left\{\bigl|\overline{Z}_{1,\mu,N}^{\rm L}-\bigl\langle{\overline{Z}_{1,\mu,N}^{\rm L}}\bigr\rangle\bigr|+\abs{\left\langle{\overline{Z}_{1,\mu,N}}\right\rangle-\overline{Z}_{1,\mu,N}}\geq{}\delta\right\}\nonumber\\
&\subseteq{}&\left\{\bigl|\overline{Z}_{1,\mu,N}^{\rm L}-\bigl\langle{\overline{Z}_{1,\mu,N}^{\rm L}}\bigr\rangle\bigr|\geq{}\frac{\delta}{2}\right\}\cup\left\{\abs{\left\langle{\overline{Z}_{1,\mu,N}}\right\rangle-\overline{Z}_{1,\mu,N}}\geq{}\frac{\delta}{2}\right\}
\end{eqnarray}
where in the first set bound we used the triangle inequality twice, in the second one we used the fact that $\bigl\langle{\overline{Z}_{1,\mu,N}^{\rm L}}\bigr\rangle-\left\langle{\overline{Z}_{1,\mu,N}}\right\rangle\leq{}0$ for all $N$ ---according to the first decoy-state bound in Eq.~(\ref{expectations})--- and in the third one we used the fact that, if $|X|+|Y|\geq{}\delta$, then either $|X|\geq{}\delta/2$ or $|Y|\geq{}\delta/2$. Now, in virtue of the union bound, we have that
\begin{equation}\label{two_terms}
P\left(E_{\delta,N}\right)\leq{}P\left(\bigl|\overline{Z}_{1,\mu,N}^{\rm L}-\bigl\langle{\overline{Z}_{1,\mu,N}^{\rm L}}\bigr\rangle\bigr|\geq{}\frac{\delta}{2}\right)+P\left(\abs{\left\langle{\overline{Z}_{1,\mu,N}}\right\rangle-\overline{Z}_{1,\mu,N}}\geq{}\frac{\delta}{2}\right).
\end{equation}
Therefore, the claim holds if we show that both terms in the right-hand side tend to zero as $N$ tends to infinity for all $\delta>0$. Recalling that, in virtue of Chebyshev's inequality~\cite{UNED}, mean-square convergence of a sequence of random variables guarantees convergence in probability, for the second term of Eq.~(\ref{two_terms}) we have
\begin{equation} \lim_{N\to\infty}Var\left[\overline{Z}_{1,\mu,N}\right]=0\implies{}\lim_{N\to\infty}P\left(\abs{\overline{Z}_{1,\mu,N}-\left\langle{\overline{Z}_{1,\mu,N}}\right\rangle}\geq{}\frac{\delta}{2}\right)=0
\end{equation}
for all $\delta>0$. Regarding the first term, note that $\overline{Z}_{1,\mu,N}^{\rm L}$ is linear in the $\overline{Z}_{a,N}$ (see Sec.~\ref{LP}). That is to say, $\overline{Z}_{1,\mu,N}^{\rm L}=\sum_{a\in{}A}c_{a}\overline{Z}_{a,N}+C$ for certain coefficients $c_{a}$ and $C$. Thus,
\begin{equation}
Var\left[\overline{Z}_{1,\mu,N}^{\rm L}\right]=E\left[\abs{\sum_{a\in{}A}c_{a}\left(\overline{Z}_{a,N}-\langle{\overline{Z}_{a,N}}\rangle\right)}^{2}\right],\\
\end{equation}
and since~\cite{UNED} $E[\abs{X+Y}^{2}]\leq{}4\bigl(E\bigl[\abs{X}^{2}\bigr]+E\bigl[\abs{Y}^{2}\bigr]\bigr)$, it follows that
\begin{equation}
Var\left[\overline{Z}_{1,\mu,N}^{\rm L}\right]\leq{}K\sum_{a}\abs{c_{a}}^{2}Var\left[\overline{Z}_{a,N}\right]
\end{equation}
for some positive constant $K$, such that
\begin{eqnarray}
&&\lim_{N\to\infty}Var\left[\overline{Z}_{a,N}\right]=0\hspace{.2cm}\textrm{for all}\hspace{.2cm}a\in{}A\implies{}\lim_{N\to\infty}Var\left[\overline{Z}_{1,\mu,N}^{\rm L}\right]=0\implies{}\nonumber\\
&&\lim_{N\to\infty}P\left(\bigl|\overline{Z}_{1,\mu,N}^{\rm L}-\bigl\langle{\overline{Z}_{1,\mu,N}^{\rm L}}\bigr\rangle\bigr|\geq{}\frac{\delta}{2}\right)=0
\end{eqnarray}
for all $\delta>0$, where again we invoked the fact that mean-square convergence implies convergence in probability.\hspace{.3cm}\qedsymbol\\

\textbf{Proposition 2.} Let us assume that $\lim_{N\to\infty}Var\left[\overline{E}_{a,N}\right]=0$ for all $a\in{}A$ and $\lim_{N\to\infty}Var\left[\overline{E}_{1,\mu,N}\right]=0$. Then, $\lim_{N\to\infty}P(\overline{E}_{1,\mu,N}\geq{}\overline{E}_{1,\mu,N}^{\rm U}+\delta)=0$ for all $\delta>0$.\\

The proof of Proposition 2 follows identically as that of Proposition 1.\\

As a final comment, note that, when dealing with bounded sequences of random variables ---$\{X_{j}\}$ is bounded if there exists some constant $C$ such that $\abs{X_{j}}<C$ for all $j$--- mean-square convergence is not stronger but exactly equivalent to convergence in probability (see for instance~\cite{UNED}), such that demanding the latter kind of convergence instead does not relax the preconditions of propositions 1 and 2. If, alternatively, neither kind of convergence is demanded, all we know is that Eq.~(\ref{expectations}) holds for the expectations, which does not suffice to establish the limits of propositions 1 and 2.%(these limits are the justification of our asymptotic formula)
%NOTE 0: The proof of equivalence between convergence in probability and mean-square convergence appears as one of the problems of my UNED book (chapter 18, problem 2?), and not in the theory part.
%NOTE 1: convergence to 0 in distribution matches exactly convergence to 0 in probability. That is why in our problem it does not make sense to comment on this third type of convergence.
%NOTE 2:A visual counterexample to justify the assertion "the positivity of the difference between expectations does not suffice to validate the asymptotic formula" is given in page 130 of my notes (oscillating observables with different amplitudes).
\section{Data availability}
No datasets were generated or analysed during the current study.
\section{Acknowledgements}
We thank Margarida Pereira for very fruitful discussions. This work was supported by the European Union's Horizon 2020 research and innovation programme under the Marie Sk\l{}odowska-Curie grant agreement No 675662 (project QCALL), by the Galician Regional Government (consolidation of Research Units: AtlantTIC), the Spanish Ministry of Economy and Competitiveness (MINECO), the Fondo Europeo de Desarrollo Regional (FEDER) through Grant No. PID2020-118178RB-C21, and the Spanish Ministry of Science and Innovation through the “Planes Complementarios de I+D+I con las Comunidades Autónomas” in Quantum Communication. V.Z. and A.N. acknowledge support from respective FPU pre-doctoral scholarships from the Spanish Ministry of Education. K.T. acknowledges support from JSPS KAKENHI Grant Numbers JP18H05237 18H05237 and JST-CREST JPMJCR 1671.
\section{Author contributions}
M.C. and K.T. conceived the initial idea and triggered the consideration of this project. V.Z. and A.N. made the theoretical analysis and performed the numerical simulations, with inputs from all authors. All authors analysed the results and prepared the manuscript.
\section{Competing interests}
The authors declare no competing interests.

\newpage
\appendix
\section{Reference values for the linearized Cauchy-Schwarz constraints}\label{reference_values}
Below, we provide the reference values $\tilde{y}_{n,a}$ and $\tilde{h}_{n,a}$ that follow from the typical channel model presented in Sec.~\ref{simulations} of the main text, which depends on the experimental inputs $\eta$, $\delta_{\rm A}$ and $p_{\rm d}$.

\begin{figure}[!htbp]
	\centering 
	\includegraphics[width=7.5cm,height=2.7cm]{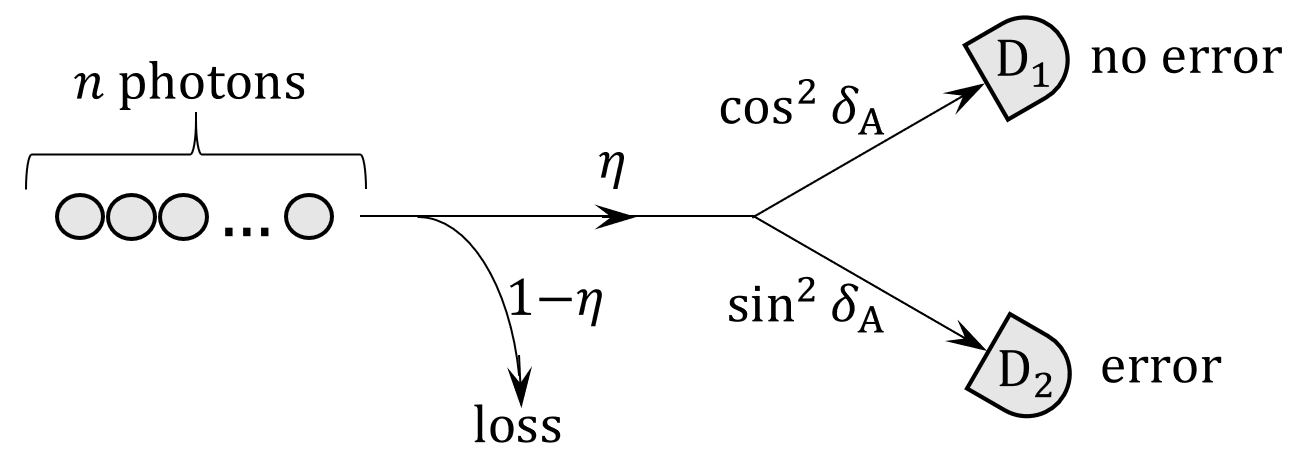}\\
	\caption{Schematic representation of the typical channel model considered in Sec.~\ref{simulations} of the main text. We recall that $n$ stands for the number of photons emitted by Alice's PRWCP source, $\eta$ stands for the overall system efficiency and $\delta_{\rm A}$ stands for the polarization misalignment.}
	\label{fig:schematic}
\end{figure}

For convenience, we calculate $\tilde{h}_{n,a}$ first, for which we proceed in two steps. Disregarding the dark counts and the random assignments of the double clicks for the moment, the possible genuine detection outcomes for an $n$-photon pulse emitted by Alice are ``no click", ``no error", ``error" and ``double click", respectively denoted as 00, 10, 01 and 11. Their probabilities are
\begin{eqnarray}\label{genuine}
&&p_{00}=(1-\eta)^{n},\nonumber \\
&&p_{10}=\left(\eta{\cos^2\delta_{\rm A}}+1-\eta\right)^{n}-(1-\eta)^{n},\nonumber \\
&&p_{01}=\left(\eta{\sin^2\delta_{\rm A}}+1-\eta\right)^{n}-(1-\eta)^{n},\nonumber \\
&&p_{11}=1-p_{00}-p_{01}-p_{10}.
\end{eqnarray}
Eq.~(\ref{genuine}) can be interpreted as follows: every photon in the $n$-photon Fock state emitted by Alice's PRWCPs source reaches Bob's lab with probability $\eta$ and experiences a polarization bit flip with probability $\sin^2\delta_{\rm A}$, as illustrated in Fig.~\ref{fig:schematic}.

In order to incorporate the dark counts and the random assignments of the double clicks, we introduce the mutually exclusive events $A=\{$no dark counts$\}$, $B=\{$dark count in $D_{1}\}$, $C=\{$dark count in $D_{2}\}$ and $D=\{$dark count in both $D_{1}$ and $D_{2}\}$, where we follow the detector notation of Fig.~\ref{fig:schematic}. The conditional error probabilities read
\begin{eqnarray}
&&p_{\rm err}\lvert_{A}=p_{01}+\frac{1}{2}p_{11},\nonumber \\
&&p_{\rm err}\lvert_{B}=\frac{1}{2}\left(p_{01}+p_{11}\right),\nonumber \\
&&p_{\rm err}\lvert_{C}=p_{00}+p_{01}+\frac{1}{2}\left(p_{10}+p_{11}\right),\nonumber\\
&&p_{\rm err}\lvert_{D}=\frac{1}{2},
\end{eqnarray}
and, consequently,
\begin{equation}\label{reference_error}
\tilde{h}_{n,a}=(1-p_{\rm d})^{2}p_{\rm err}\lvert_{A}+p_{\rm d}(1-p_{\rm d})\left(p_{\rm err}\lvert_{B}+p_{\rm err}\lvert_{C}\right)+p_{\rm d}^{2}p_{\rm err}\lvert_{D}
\end{equation}
for all $n\in\mathbb{N}$ and $a\in{}A$. Of course, regarding $\tilde{y}_{n,a}$, we have
\begin{equation}\label{reference_yield}
\tilde{y}_{n,a}=1-(1-p_{\rm d})^{2}p_{00}
\end{equation}
for all $n\in\mathbb{N}$ and $a\in{}A$.
\section{Trace distance argument}\label{TD}
The trace distance (TD) argument is stated as follows.\\

\textbf{Theorem}~\cite{Nielsen}. Let $\rho$ and $\sigma$ be two distinct states of a certain quantum system. Then, the trace distance between $\rho$ and $\sigma$ satisfies $D(\rho,\sigma)=\max\{\mathrm{Tr}(\hat{O}(\rho-\sigma))\}$, where the maximization is taken over all positive operators $\hat{O}\leq{I}$.

\begin{figure}[!htbp]
	\centering
	\includegraphics[width=9.6cm,height=8.1cm]{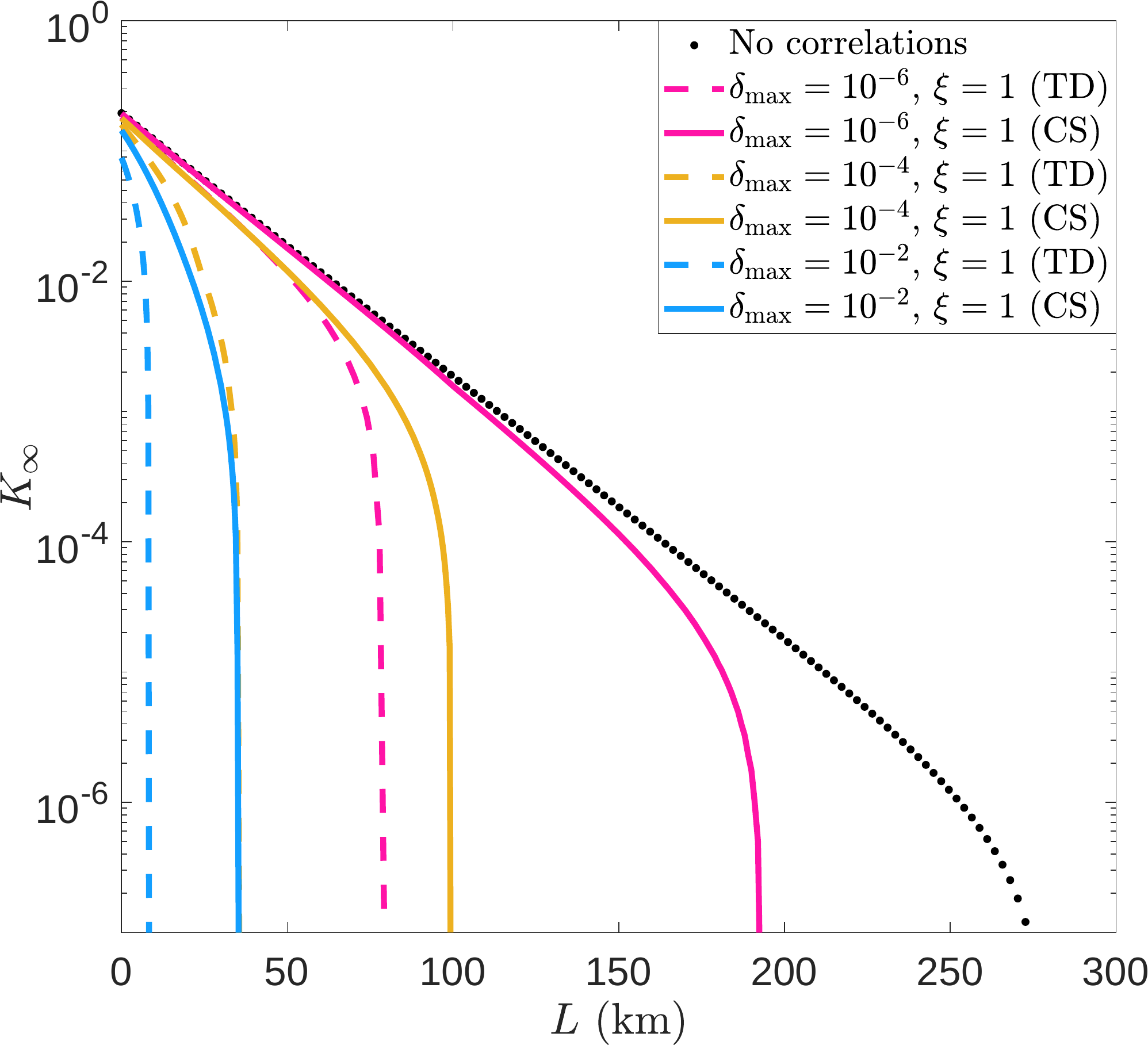}\\
	\caption{Comparison between the TD argument and the linearized CS constraint in terms of their secret key rate performance. For illustration purposes, only nearest-neighbors intensity correlations are contemplated, \textit{i.e.,} we set $\xi=1$. On the one hand, the dashed lines are obtained using the TD argument, showing the asymptotic secret key rate, $K_{\infty}$, as a function of the distance, $L$, for various values of the maximum relative deviation between intensity settings ($a_{k}$) and actual intensities ($\alpha_{k}$), $\delta_{\rm max}\in\{10^{-6},10^{-4},10^{-2}\}$. On the other hand, the solid lines represent the corresponding secret key rates obtained with the CS inequality instead. Although these latter lines also appear in Fig.~\ref{fig:model-independent} of the main text, for clarity purposes the color and line style criteria are different here. In addition, we include the attainable secret key rate in the absence of intensity correlations for completeness (dotted black line). Regarding the experimental parameters, they are fixed identically as in Fig.~\ref{fig:model-independent} of the main text.}
	\label{fig:model-independent_TD}
\end{figure}

Keeping the notation $Y_{n,a}^{(k)}=p^{(k)}(\mathrm{click}|n,a,\mathrm{Z,Z})$, and making use of the fact that $D(\ketbra{x}{x},\ketbra{y}{y})=(1-\abs{\braket{x}{y}}^{2})^{1/2}$~\cite{Nielsen}, the bound provided by the TD argument reads~\cite{Tamaki}
\begin{equation}\label{TD_deviation}
\left\lvert{Y_{n,a}^{(k)}-Y_{n,b}^{(k)}}\right\rvert\leq{}\sqrt{1-\abs{\braket{{\Psi}^{(k)}_{b,\mathrm{Z},n}}{{\Psi}^{(k)}_{a,\mathrm{Z},n}}}^{2}}\leq{}\sqrt{1-\tau_{ab,n}^{\xi}}
\end{equation}
for all $a\in{}A$, $b\in{}A$ ($b\neq{}a$), $n\in\mathbb{N}$ and $k=1,\ldots,N$. Here, we have used the lower bound
\begin{equation} \abs{\braket{{\Psi}^{(k)}_{b,\mathrm{Z},n}}{{\Psi}^{(k)}_{a,\mathrm{Z},n}}}^{2}\geq{}\tau_{ab,n}^{\xi}
\end{equation}
presented in the main text, which depends on a presumed finite correlation range $\xi$. Remarkably, Eq.~(\ref{TD_deviation}) does not rely on a characterization of the quantum channel, as opposed to the linearized CS constraints. What is more, the TD argument provides equivalent constraints for the $n$-photon error click probabilities too, as seen next. In the first place,
\begin{equation}
\left\lvert{H_{n,a,r}^{(k)}-H_{n,b,r}^{(k)}}\right\rvert\leq{}\sqrt{1-\abs{\braket{{\Psi}^{(k)}_{b,\mathrm{X},r,n}}{{\Psi}^{(k)}_{a,\mathrm{X},r,n}}}^{2}}
\end{equation}
for $n\in{}\mathbb{N}$, $a\in{}A$, $b\in{}A$ ($b\neq{}a$), $r\in\mathbb{Z}_{2}$ and $k=1,\ldots,N$, where we maintain the notation $H_{n,a,r}^{(k)}=p^{(k)}(\mathrm{err}|n,a,\mathrm{X},\mathrm{X},r)$ and $H_{n,a}^{(k)}=p^{(k)}(\mathrm{err}|n,a,\mathrm{X,X})$. If, in addition, we recall that $\abs*{\braket*{{\Psi}^{(k)}_{b,\mathrm{X},r,n}}{{\Psi}^{(k)}_{a,\mathrm{X},r,n}}}=\abs*{\braket*{{\Psi}^{(k)}_{b,\mathrm{Z},n}}{{\Psi}^{(k)}_{a,\mathrm{Z},n}}}$ for both $r=0$ and $r=1$, the desired bound follows from the triangle inequality:
\begin{eqnarray}\label{triangle}
&&\left\lvert{H_{n,a}^{(k)}-H_{n,b}^{(k)}}\right\rvert=\left\lvert{\frac{1}{2}\left(H_{n,a,0}^{(k)}+H_{n,a,1}^{(k)}\right)-\frac{1}{2}\left(H_{n,b,0}^{(k)}+H_{n,b,1}^{(k)}\right)}\right\rvert\leq{}\nonumber \\
&&\frac{1}{2}\left\lvert{H_{n,a,0}^{(k)}-H_{n,b,0}^{(k)}}\right\rvert+\frac{1}{2}\left\lvert{H_{n,a,1}^{(k)}-H_{n,b,1}^{(k)}}\right\rvert\leq{}\sqrt{1-\tau_{ab,n}^{\xi}}
\end{eqnarray}
for any given finite correlation range $\xi$.

Aiming to compare the TD argument and the linearized CS constraints in terms of their secret key rate performance, one must replace the corresponding restrictions by Eq.~(\ref{TD_deviation}) and Eq.~(\ref{triangle}) in the linear programs of Sec.~\ref{LP}. The result is illustrated in Fig.~\ref{fig:model-independent_TD} for the most representative case of nearest-neighbors intensity correlations, \textit{i.e.,} $\xi=1$.

Comparing Fig.~\ref{fig:model-independent_TD} with Fig.~\ref{fig:model-independent} in the main text, we see that the linearized CS constraint provides significantly tighter bounds than the TD argument for the parameter estimation, as long as adequate reference parameters are given as inputs to the former.

\section{Deterministic intensity correlations model}\label{deterministic}
In this note, we consider a deterministic intensity correlations model where, at every round $k$, the record of settings ($\vec{a}_{k}$) fully determines the intensity ($\alpha_{k}$), instead of just pinning its probability distribution. Nevertheless, we assume that the exact value of $\alpha_{k}$ is unknown to keep the analysis as general as possible. That is to say, for any given record $\vec{a}_{k}$, the model reads
\begin{equation}\label{deterministic_model}
g_{\vec{a}_{k}}(\alpha_{k})=\delta\left(\alpha_{k}-\alpha_{k}^{\vec{a}_{k}}\right),
\end{equation}
for some unknown $\alpha_{k}^{\vec{a}_{k}}\in[a_{k}^{-},a_{k}^{+}]$ fixed by $\vec{a}_{k}$ (the worst case will be considered), where $\delta(\cdot)$ stands for the Dirac delta distribution. This model allows to compute a tighter lower bound for the overlap $\abs*{\braket*{{\Psi}^{(k)}_{b,\mathrm{Z},n}}{{\Psi}^{(k)}_{a,\mathrm{Z},n}}}$ than the one derived in the model-independent case. For this purpose, the starting point is the equation
\begin{eqnarray}\label{overlap_finite_range_repeat}
&&\braket{{\Psi}^{(k)}_{b,\mathrm{Z},n}}{{\Psi}^{(k)}_{a,\mathrm{Z},n}}=\nonumber\\
&&\sum_{a_{\max\{k-\xi,1\}}^{k-1}}\sqrt{p^{(k)}({a}_{k-1},\ldots,{a}_{\max\{k-\xi,1\}}|n,a,\mathrm{Z})p^{(k)}({a}_{k-1},\ldots,{a}_{\max\{k-\xi,1\}}|n,b,\mathrm{Z})}\nonumber \\
&&\times\left[\sum_{a_{k+1}^{\min\{k+\xi,N\}}}\left(\prod_{i=k+1}^{\min\{k+\xi,N\}}p_{a_{i}}\braket{\psi_{\vec{a}_{i}(a_{k}=b)}}{\psi_{\vec{a}_{i}(a_{k}=a)}}_{B_{i}C_{i}}\right)\right]
\end{eqnarray}
for $k=2,\ldots,N-1$, where we recall that
\begin{equation}\label{un-referenced_4}
\braket{\psi_{\vec{a}_{i}(a_{k}=b)}}{\psi_{\vec{a}_{i}(a_{k}=a)}}_{B_{i}C_{i}}=\sum_{n=0}^{\infty}\left(p_{n}\lvert_{\vec{a}_{i}(a_{k}=b)}\times{}p_{n}\lvert_{\vec{a}_{i}(a_{k}=a)}\right)^{1/2}
\end{equation}
for all $i=k+1,\ldots,N$, and $\xi$ stands for the finite correlation range. Noticing that Eq.~(\ref{deterministic_model}) implies
\begin{equation} p_{n}\lvert_{\vec{a}_{i}(a_{k}=b)}=\frac{\exp{-\alpha_{i}^{\vec{a}_{i}(a_{k}=b)}}\left(\alpha_{i}^{\vec{a}_{i}(a_{k}=b)}\right)^{n}}{n!}
\end{equation}
for a fixed (but unknown) $\alpha_{i}^{\vec{a}_{i}(a_{k}=b)}\in[a_{i}^{-},a_{i}^{+}]$, $a_{i}\in{}A$, it follows that
\begin{eqnarray} &&\braket{\psi_{\vec{a}_{i}(a_{k}=b)}}{\psi_{\vec{a}_{i}(a_{k}=a)}}_{B_{i}C_{i}}=\nonumber\\
&&\sum_{n=0}^{\infty}\exp\left\{-\left(\alpha_{i}^{\vec{a}_{i}(a_{k}=a)}+\alpha_{i}^{\vec{a}_{i}(a_{k}=b)}\right)/2\right\}\sqrt{\alpha_{i}^{\vec{a}_{i}(a_{k}=a)}\alpha_{i}^{\vec{a}_{i}(a_{k}=b)}}^{\hspace{.05cm}n}/n!=\nonumber \\
&&\exp\left\{\sqrt{\alpha_{i}^{\vec{a}_{i}(a_{k}=a)}\alpha_{i}^{\vec{a}_{i}(a_{k}=b)}}-\left(\alpha_{i}^{\vec{a}_{i}(a_{k}=a)}+\alpha_{i}^{\vec{a}_{i}(a_{k}=b)}\right)/2\right\}.
\end{eqnarray}
Now, analytically minimizing this overlap for $\left(\alpha_{i}^{\vec{a}_{i}(a_{k}=a)},\alpha_{i}^{\vec{a}_{i}(a_{k}=b)}\right)\in[a_{i}^{-},a_{i}^{+}]\times[a_{i}^{-},a_{i}^{+}]$, one obtains
\begin{equation} \braket{\psi_{\vec{a}_{i}(a_{k}=b)}}{\psi_{\vec{a}_{i}(a_{k}=a)}}_{B_{i}C_{i}}\geq{}\exp\left\{\sqrt{a_{i}^{+}a_{i}^{-}}-(a_{i}^{+}+a_{i}^{-})/2\right\},
\end{equation}
such that the bracket in Eq.~(\ref{overlap_finite_range_repeat}) is lower-bounded as
\begin{eqnarray}
&&\sum_{a_{k+1}^{\min\{k+\xi,N\}}}\left(\prod_{i=k+1}^{\min\{k+\xi,N\}}p_{a_{i}}\braket{\psi_{\vec{a}_{i}(a_{k}=b)}}{\psi_{\vec{a}_{i}(a_{k}=a)}}_{B_{i}C_{i}}\right)\geq{}\nonumber\\
&&\left[\sum_{c\in{}A}p_{c}\exp\left\{\sqrt{c^{+}c^{-}}-(c^{+}+c^{-})/2\right\}\right]^{\xi},
\end{eqnarray}
which factors off the remaining summations of Eq.~(\ref{overlap_finite_range_repeat}). For the latter, we maintain the model-independent bound provided in the main text for simplicity. Thus, putting both terms together and recalling that the resulting bound also applies to the extreme rounds $k=1$ and $k=N$ (this consequence follows identically as in the Methods Sec.~\ref{Cauchy}), we conclude that
\begin{eqnarray}\label{gamma}
&&\abs{\braket{{\Psi}^{(k)}_{b,\mathrm{Z},n}}{{\Psi}^{(k)}_{a,\mathrm{Z},n}}}^{2}\geq{}\nonumber\\
&&\gamma_{ab,n}^{\xi}= \begin{cases}
e^{a^{-}+b^{-}-(a^{+}+b^{+})}\left[\displaystyle\sum_{c\in{}A}p_{c}\exp\left\{\sqrt{c^{+}c^{-}}-(c^{+}+c^{-})/2\right\}\right]^{2\xi} & \mathrm{if}\hspace{.3cm}n=0\nonumber\\
e^{a^{+}+b^{+}-(a^{-}+b^{-})}\left(\frac{a^{-}b^{-}}{a^{+}b^{+}}\right)^{n}\left[\displaystyle\sum_{c\in{}A}p_{c}\exp\left\{\sqrt{c^{+}c^{-}}-(c^{+}+c^{-})/2\right\}\right]^{2\xi} & \mathrm{if}\hspace{.3cm}n\geq{}1\nonumber\\
\end{cases}\\
\end{eqnarray}
for all $k=1,\ldots{}N$, $n\in{}\mathbb{N}$, $a\in{}A$, $b\in{}A$ and $b\neq{}a$.

Remarkably, the only difference introduced by the deterministic model in the parameter estimation procedure consists of replacing $\tau_{ab,n}^{\xi}$ by $\gamma_{ab,n}^{\xi}$ everywhere in the linearized CS constraints. Beyond this, the linear programs of Sec.~\ref{LP} remain unchanged.
\clearpage
\begin{figure}[t!]%[!htbp]
	\centering
	\includegraphics[width=9.6cm,height=8.1cm]{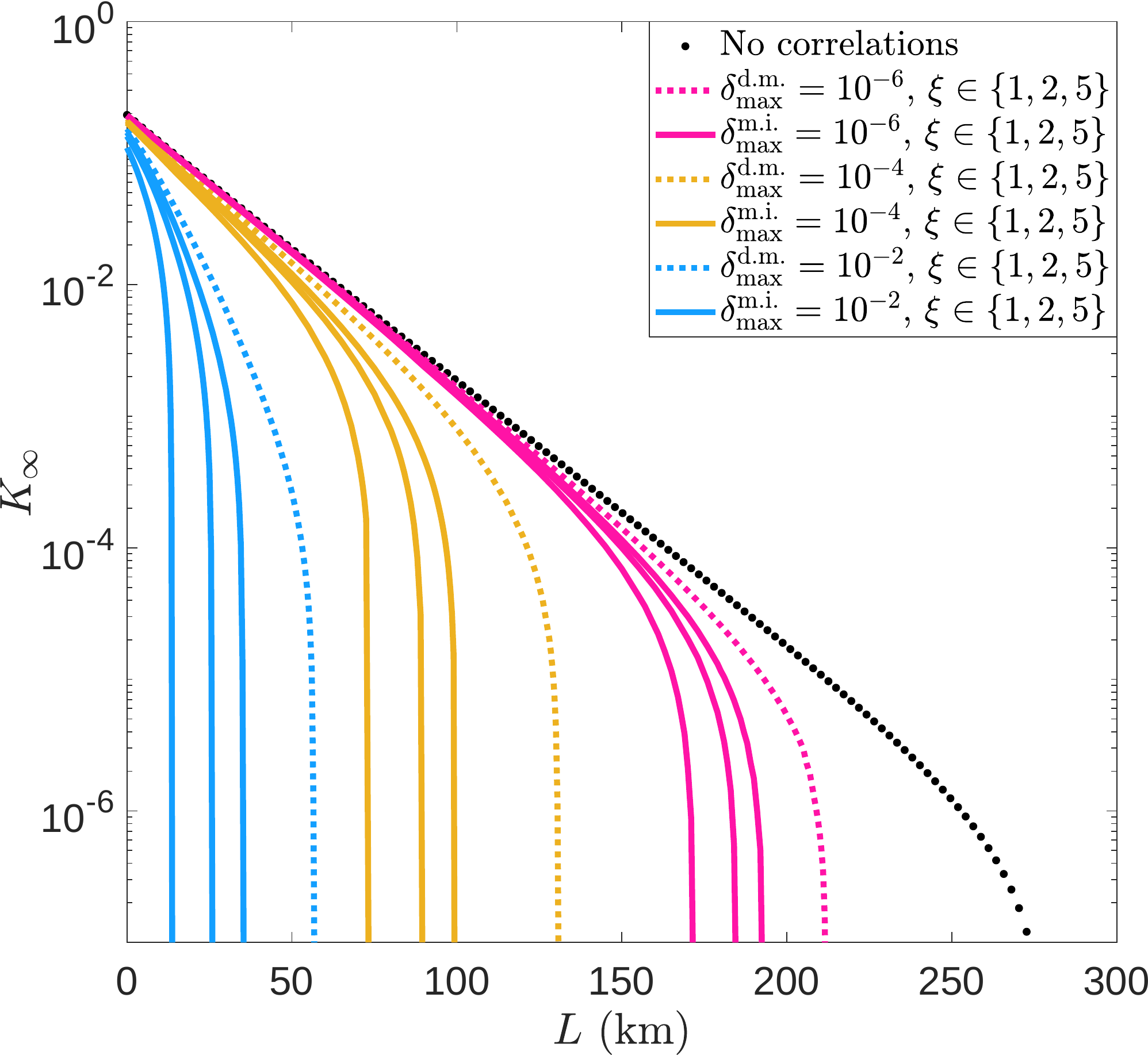}\\
	\caption{Comparison between deterministic intensity correlations (denoted by ``d.m." in the figure legend) and model-independent intensity correlations (denoted by ``m.i." in the figure legend) in terms of their secret key rate performance. The dotted color lines show the asymptotic secret key rate, $K_{\infty}$, as a function of the distance, $L$, assuming the deterministic model. In accordance with Fig.~\ref{fig:model-independent_TD}, we contemplate three different values of the maximum relative deviation between intensity settings ($a_{k}$) and actual intensities ($\alpha_{k}$), $\delta_{\rm max}\in\{10^{-6},10^{-4},10^{-2}\}$. Similarly, the solid color lines represent the corresponding secret key rates in the model-independent scenario. Importantly, three correlation ranges are used, namely, $\xi\in\{1,2,5\}$. However, the effect of this parameter on the secret key rate is negligible in the deterministic model for such moderate values, and thus we only plot $\xi=1$ in that case. As a reference, we provide the attainable key rate in the absence of intensity correlations too (black dotted line). Regarding the experimental parameters, they are common with those of Fig.~\ref{fig:model-independent} in the main text.}
	\label{fig:deterministic}
\end{figure}

To finish with, Fig.~\ref{fig:deterministic} illustrates how the asymptotic secret key rate $K_{\infty}$ is enhanced when one moves from the model-independent scenario of Fig.~\ref{fig:model-independent} in the main text to a deterministic model. Remarkably, as seen in the figure, this model is also more robust to the correlation range $\xi$ than the model-independent setting.

\begin{thebibliography}{40}
\bibitem{Scarani}
Scarani, V. \textit{et al}.
\href{https://doi.org/10.1103/RevModPhys.81.1301}{The security of practical quantum key distribution. \textit{Rev. Mod. Phys.} \textbf{81,} 1301 (2009).}
%
\bibitem{Curty}
Lo, H.-K., Curty, M. \& Tamaki, K. \href{https://doi.org/10.1038/nphoton.2014.149}{Secure quantum key distribution. \textit{Nat. Photonics} \textbf{8,} 595 (2014).}
%
\bibitem{Feihu}
Xu, F., Ma, X., Zhang, Q., Lo, H.-K., \& Pan, J.-W. \href{https://doi.org/10.1103/RevModPhys.92.025002}{Secure quantum key distribution with realistic devices. \textit{Rev. Mod. Phys.} \textbf{92,} 025002 (2020).}
%
\bibitem{OTP}
Vernam, G. S., \textit{Trans. Am. Inst. Electr. Eng.} XLV, 295 (1926).
\bibitem{BB84}
Bennett, C. H. \& Brassard, G. Quantum cryptography: public key distribution and coin tossing. \textit{In Proc. IEEE International Conference on Computers, Systems \& Signal Processing}, 175–179 (IEEE, NY, Bangalore, India, 1984).
%
\bibitem{Pan1}
Yin, H. L., \textit{et al}. \href{https://doi.org/10.1103/PhysRevLett.117.190501}{Measurement-device-independent quantum key distribution over a 404 km optical fiber. \textit{Phys. Rev. Lett.} \textbf{117,} 190501 (2016).}
%
\bibitem{Boaron}
Boaron, A., \textit{et al}. \href{https://doi.org/10.1103/PhysRevLett.121.190502}{Secure quantum key distribution over 421 km of optical fiber. \textit{Phys. Rev. Lett.} \textbf{121,} 190502 (2018).}
%
\bibitem{Pan2}
Fang, X. T., \textit{et al}. \href{https://doi.org/10.1038/s41566-020-0599-8}{Implementation of quantum key distribution surpassing the linear rate-transmittance bound. \textit{Nat. Photonics} \textbf{14,} 422-425 (2020).}
%
\bibitem{Pan3}
Chen, J. P., \textit{et al}. \href{https://doi.org/10.1103/PhysRevLett.124.070501}{Sending-or-not-sending with independent lasers: secure twin-field quantum key distribution over 509 km. \textit{Phys. Rev. Lett.} \textbf{124,} 070501 (2020).}
%
\bibitem{Tomita}
Yoshino, K. I. \href{https://doi.org/10.1038/s41534-017-0057-8}{\textit{et al}. Quantum key distribution with an efficient countermeasure against correlated intensity fluctuations in optical pulses. \textit{npj Quantum Inf.} \textbf{4,} 1-8 (2018).}
%
\bibitem{Geneve}
Grünenfelder, F., Boaron, A., Rusca, D., Martin, A. \& Zbinden, H. \href{https://doi.org/10.1063/5.0021468}{Performance and security of 5 GHz repetition rate polarization-based quantum key distribution. \textit{Appl. Phys. Lett.} \textbf{117,} 144003 (2020).}
%
\bibitem{decoy2}
Hwang, W. Y. \href{https://doi.org/10.1103/PhysRevLett.91.057901}{Quantum key distribution with high loss: toward global secure communication. \textit{Phys. Rev. Lett.} \textbf{91,} 057901 (2003).}
%
\bibitem{decoy}
Lo, H.-K., Ma, X. \& Chen, K. \href{https://doi.org/10.1103/PhysRevLett.94.230504}{Decoy state quantum key distribution. \textit{Phys. Rev. Lett.} \textbf{94,} 230504 (2005).}
%
\bibitem{decoy3}
Wang, X.-B. \href{https://doi.org/10.1103/PhysRevLett.94.230503}{Beating the photon-number-splitting attack in practical quantum cryptography. \textit{Phys. Rev. Lett.} \textbf{94,} 230503 (2005).}
%
\bibitem{Lim}
Lim, C. C. W., Curty, M., Walenta, N., Xu, F. \& Zbinden, H. \href{https://doi.org/10.1103/PhysRevA.89.022307}{Concise security bounds for practical decoy-state quantum key distribution. \textit{Phys. Rev. A} \textbf{89,} 022307 (2014).}
%
\bibitem{Tamaki}
Tamaki, K., Curty, M., \& Lucamarini, M. \href{https://doi.org/10.1088/1367-2630/18/6/065008}{Decoy-state quantum key distribution with a leaky source. \textit{New J. Phys.} \textbf{18,} 065008 (2016).}
%
\bibitem{Nagamatsu}
Nagamatsu Y., Mizutani, A., Ikuta, R., Yamamoto, T., Imoto, N., \& Tamaki, K. \href{https://doi.org/10.1103/PhysRevA.93.042325}{Security of quantum key distribution with light sources that are not independently and identically distributed. \textit{Phys. Rev. A} \textbf{93,} 042325 (2016).}
%
\bibitem{Mizutani}
Mizutani, A. \textit{et al}. \href{https://doi.org/10.1038/s41534-018-0122-y}{Quantum key distribution with setting-choice-independently correlated light sources. \textit{npj Quantum Inf.} \textbf{5,} 8 (2019).}
%
\bibitem{Roberts}
Roberts, G. L. \textit{et al.} \href{https://doi.org/10.1364/OL.43.005110}{Patterning-effect mitigating intensity modulator for secure decoy-state quantum key distribution. \textit{Optics letters} \textbf{43,} 5110-5113 (2018).}
%
\bibitem{Pereira}
Pereira, M., Kato, G., Mizutani, A., Curty, M. \& Tamaki, K. \href{https://doi.org/10.1126/sciadv.aaz4487}{Quantum key distribution with correlated sources. \textit{Science Advances} \textbf{6,} eaaz4487  (2020).}
%
\bibitem{MDI1}
Lo, H.-K., Curty, M., \& Qi, B. \href{https://doi.org/10.1103/PhysRevLett.108.130503}{Measurement-device-independent quantum key distribution. \textit{Phys. Rev. Lett.} \textbf{108,} 130503 (2012).}
%
\bibitem{TF}
Lucamarini, M., Yuan, Z., Dynes, J., \& Shields, A. \href{https://doi.org/10.1038/s41586-018-0066-6}{Overcoming the rate–distance limit of quantum key distribution without quantum repeaters. \textit{Nature} \textbf{557,} 400 (2018).}
%
\bibitem{Lo-Preskill}
Lo, H.-K. \& Preskill, J. \href{https://doi.org/10.26421/QIC7.5-6-2}{Security of quantum key distribution using weak coherent states with nonrandom phases. \textit{Quantum Inf. Comput.} \textbf{7,} 431–458 (2007).}
%
\bibitem{Chernoff}
Mitzenmacher, M., \& Upfal, E. Probability and computing: Randomization and probabilistic techniques in algorithms and data analysis (Cambridge University Press, 2017).
%
\bibitem{Hoeffding}
Hoeffding, W. Probability inequalities for sums of bounded random variables. \textit{J. Am. Stat. Assoc.} \textbf{58,} 13-30 (1963).
%
\bibitem{Zapatero}
Zapatero, V., \& Curty, M. \href{https://doi.org/10.1038/s41534-020-00358-y}{Secure quantum key distribution with a subset of malicious devices. \textit{npj Quantum Inf.} 7, 1-8 (2021).}
%
\bibitem{Nielsen}
Nielsen, M. \& Chuang, I. \textit{Quantum Computation and Quantum Information}, Cambridge University Press (2000).
%
\bibitem{Navarrete}
Navarrete, Á., Pereira, M., Curty, M., \& Tamaki K. \href{https://doi.org/10.1103/PhysRevApplied.15.034072}{Practical quantum key distribution that is secure against side channels. \textit{Phys. Rev. Appl.} \textbf{15,} 034072 (2021).}
%
\bibitem{LP}
%Eduardo Ramos Méndez, \textit{Programación Lineal y Entera}, Sanz y Torres (2020).
Bazaraa, M. S., Jarvis, J. J., \& Sherali, H. D. \href{https://doi.org/10.1002/9780471703778}{\textit{Linear programming and network flows}, John Wiley \& Sons (2008).}
%
\bibitem{Tomamichel}
Tomamichel, M., Lim, C. C. W., Gisin, N., \& Renner, R. \href{https://doi.org/10.1038/ncomms1631}{Tight finite-key analysis for quantum cryptography. \textit{Nat. Commun.} \textbf{3,} 1-6 (2012).}
%
\bibitem{Serfling}
Serfling, R.~J. \href{https://doi.org/10.1214/aos/1176342611}{Probability inequalities for the sum in sampling without replacement. \textit{Ann. Stat.}, 39-48 (1974).}
%
\bibitem{UNED}
%Ricardo Vélez Ibarrola, \textit{Cálculo de Probabilidades 2}, UNED (2019).
Billingsley, P. \href{https://doi.org/10.1002/9780470316962}{\textit{Convergence of probability measures}, John Wiley \& Sons (2013).}
\end{thebibliography}
\end{document}